\documentclass[%
aip,
rsi,%
amsmath,
amssymb,
reprint,%
floatfix
]{revtex4-1}

\usepackage{graphicx}
\usepackage{dcolumn}
\usepackage{bm}

\usepackage{xcolor}

\begin{document}

\title[Blue microlasers integrated on a photonic platform on silicon]{Blue microlasers integrated on a photonic platform on silicon}

\author{Farsane Tabataba-Vakili}
\affiliation{Centre de Nanosciences et de Nanotechnologies, CNRS, Univ. Paris-Sud, Universit\'{e} Paris-Saclay, F-91405 Orsay, France.}
\affiliation{CEA, INAC-PHELIQS, Nanophysique et semiconducteurs group, Univ. Grenoble Alpes, F-38000 Grenoble, France.}
\author{Laetitia Doyennette}
\affiliation{Laboratoire Charles Coulomb (L2C), Universit\'e de Montpellier, CNRS, F-34095 Montpellier, France.}
\author{Christelle Brimont}
\affiliation{Laboratoire Charles Coulomb (L2C), Universit\'e de Montpellier, CNRS, F-34095 Montpellier, France.}
\author{Thierry Guillet}
\affiliation{Laboratoire Charles Coulomb (L2C), Universit\'e de Montpellier, CNRS, F-34095 Montpellier, France.}
\author{St\'{e}phanie Rennesson}
\affiliation{Universit\'{e} C\^{o}te d'Azur, CRHEA-CNRS, F-06560 Valbonne, France.}
\author{Eric Frayssinet}
\affiliation{Universit\'{e} C\^{o}te d'Azur, CRHEA-CNRS, F-06560 Valbonne, France.}
\author{Benjamin Damilano}
\affiliation{Universit\'{e} C\^{o}te d'Azur, CRHEA-CNRS, F-06560 Valbonne, France.}
\author{Jean-Yves Duboz}
\affiliation{Universit\'{e} C\^{o}te d'Azur, CRHEA-CNRS, F-06560 Valbonne, France.}
\author{Fabrice Semond}
\affiliation{Universit\'{e} C\^{o}te d'Azur, CRHEA-CNRS, F-06560 Valbonne, France.}
\author{Iannis Roland}
\altaffiliation{Current address: Universit\'e Paris Diderot-Paris7, F-75013 Paris, France.}
\affiliation{Centre de Nanosciences et de Nanotechnologies, CNRS, Univ. Paris-Sud, Universit\'{e} Paris-Saclay, F-91405 Orsay, France.}
\author{Moustafa El Kurdi}
\affiliation{Centre de Nanosciences et de Nanotechnologies, CNRS, Univ. Paris-Sud, Universit\'{e} Paris-Saclay, F-91405 Orsay, France.}
\author{Xavier Checoury}
\affiliation{Centre de Nanosciences et de Nanotechnologies, CNRS, Univ. Paris-Sud, Universit\'{e} Paris-Saclay, F-91405 Orsay, France.}
\author{S\'ebastien Sauvage}
\affiliation{Centre de Nanosciences et de Nanotechnologies, CNRS, Univ. Paris-Sud, Universit\'{e} Paris-Saclay, F-91405 Orsay, France.}
\author{Bruno Gayral}
\affiliation{CEA, INAC-PHELIQS, Nanophysique et semiconducteurs group, Univ. Grenoble Alpes, F-38000 Grenoble, France.}
\author{Philippe Boucaud}
\email{philippe.boucaud@crhea.cnrs.fr}
\affiliation{Centre de Nanosciences et de Nanotechnologies, CNRS, Univ. Paris-Sud, Universit\'{e} Paris-Saclay, F-91405 Orsay, France.}
\affiliation{Universit\'{e} C\^{o}te d'Azur, CRHEA-CNRS, F-06560 Valbonne, France.}

\begin{abstract}
The main interest of group-III nitride nanophotonic circuits is the integration of active structures and laser sources. A photonic platform of group-III nitride microdisk lasers integrated on silicon (Si) and emitting in the blue spectral range is demonstrated. The active microdisks are side-coupled to suspended bus waveguides and the coupled emission is guided and out-coupled to free space using grating couplers. A small gap size of less than 100 nm between the disk and the waveguide is required in the blue spectral range for optimal evanescent coupling. To avoid reabsorption of the microdisk emission in the waveguide, the quantum wells (QWs) are etched away from the waveguide. Under continuous-wave (cw) excitation, loaded quality (Q) factors greater than 2000 are observed for the whispering gallery modes (WGMs) for devices with small gaps and large waveguide bending angles. Under pulsed excitation conditions, lasing is evidenced for $3 ~\mu \text{m}$ diameter microdisks integrated in a full photonic circuit. We thus present a first demonstration of a III-nitride microlaser coupled to a nanophotonic circuit.
\end{abstract}

\maketitle

\section{Introduction}
In recent years, group-III nitride on silicon nanophotonics has been a field of considerable interest as this material system poses a promising platform for photonics. The first asset of III-nitrides is the large transparency window from the near-infrared to the blue and ultra-violet (UV) spectral range\cite{Xiong2012,Pernice2012,Neel2014,Stegmaier2014,Roland2016}. The monolithic integration of active emitters at short wavelength is certainly the most striking second advantage as compared to other photonic platforms. An integrated photonic platform like the one based on indium phosphide (InP) is limited to the near-infrared spectral range \cite{Smit2014}. Both silicon (Si) and silicon nitride (SiN) photonics are lacking efficient integrated active laser emitters. The third advantage of the III-nitride materials for photonics is their possible co-integration with III-nitride electronics, thus offering a complete toolbox for designers. The application domains of III-nitride photonics cover quantum technologies, including emission of entangled photon pairs in the near-infrared \cite{Guo2017}, the optical manipulation of ion-trapped qubits in the UV \cite{Leonardis2017} and quantum sensing \cite{Grosso2017,Berhane2017}. Bio-photonic applications in the blue and visible spectral ranges, gallium nitride (GaN) being bio-compatible \cite{Jewett2012,Hofstetter2012}, and visible communications with micro-displays and light fidelity (Li-Fi) communications \cite{Chi2015} might as well benefit from this platform.

So far, photonic circuits using light emitting diodes (LEDs)\cite{Shi2017} and passive photonic circuits using sputtered aluminum nitride (AlN) on oxide \cite{Pernice2012,Pernice20122,Xiong20122,Stegmaier2014} have been demonstrated on silicon substrates. In parallel, there have been numerous reports on single III-nitride microdisk resonators and lasers on silicon and sapphire substrates with Q factors larger than $10,000$ \cite{Tamboli2007,Simeonov2007,Simeonov2008,Mexis2011,Aharonovich2013,Athanasiou2014,Zhang2015,Selles2016,Rousseau2018}. Here we demonstrate the combination of a monolithic blue microlaser emitter with photonic circuitry on silicon.  
A major difficulty in combining a microdisk laser containing quantum wells (QWs) as the active medium with a waveguide is to avoid re-absorption of the emission in the waveguide. We are tackling this issue by partially etching the waveguide to remove the QWs, which are within 120 nm of the surface. Similar active devices have been fabricated in other material systems with absorbing waveguides \cite{Koseki2009,Schmidt2017}, while selective regrowth has also been proposed as a method to avoid re-absorption \cite{Matsuo2010}. Another technological challenge is achieving the small gap size of less than 100 nm between the disk and the waveguide that is required in the blue spectral range for efficient coupling. Furthermore, the III-nitride photonic circuit must be suspended to avoid absorption in the silicon substrate and to have a good vertical confinement by refractive index contrast to air. We have successfully fabricated suspended active III-nitride photonic circuits containing a microdisk laser, a bus waveguide with a gap size as small as 80 nm and out-coupling gratings at both ends of the waveguide (see figure \ref{fig:sem}). For waveguides with a large bending angle around the disk and small gap sizes, we observe large loaded quality factors greater than 2000 under low-power cw pumping. For microdisks with larger gaps and straight waveguides, we have observed lasing under pulsed optical pumping conditions from the disk's scattered light and from guided light outcoupled to free space through the gratings.

\section{Results and discussion}

The investigated sample was grown on a 2 inch silicon (111) substrate using molecular beam epitaxy (MBE). First a buffer layer consisting of 100 nm AlN, 100 nm of n-doped GaN (silicon concentration of $5\cdot 10^{19}~\text{cm}^{-3}$), and 200 nm of undoped GaN was deposited. Then the active region was grown consisting of 10 pairs of 2.2 nm 12\% indium gallium nitride ($\text{In}_{0.12}\text{Ga}_{0.88}\text{N}$) QWs and 9 nm GaN barriers.

We hereafter present our technological approach to couple a III-nitride microlaser to an integrated photonic circuit on a silicon platform. Microdisk photonic circuits were fabricated using three levels of electron beam lithography with UV3 resist (diluted 1:1 with EC solvent) and inductively coupled plasma (ICP) etching using boron trichloride ($\text{BCl}_3$) and chlorine ($\text{Cl}_2$) gases. For each level, plasma enhanced chemical vapor deposition (PECVD) silicon dioxide ($\text{SiO}_2$) was used as a hard mask during ICP etching. In the first level the microdisk and bus waveguide are defined and the surrounding area is etched to the substrate (a non-continuous thin AlN layer remains around the microdisks). The QWs are removed selectively from the waveguide in the second level by opening the area containing the previously defined waveguide and etching to a depth of 120 nm. In the third level, the gratings are defined at both ends of the waveguide and etched 200 nm deep. Subsequently, the devices are isotropically underetched by xenon difluoride ($\text{XeF}_2$) in principal leaving the disk with a silicon pedestal and the waveguide and grating suspended. Some of the $3~\mu \text{m}$ disks are suspended without a pedestal through a thin layer of AlN at the foot of the disk. The total device length is $60~\mu \text{m}$ with about $20~\mu \text{m}$ between the edge of the disk and the beginning of the grating. The microdisks are $3~\mu \text{m}$ and $5~\mu \text{m}$ in diameter. The bending angle of the waveguide around the disk has been varied between $0^\circ$ and $90^\circ$, where $0^\circ$ corresponds to a straight waveguide and $90^\circ$ corresponds to the waveguide being bent around one quarter of the disk's circumference. The gap size $g$ was varied between 80 nm and 120 nm and the grating period between 170 nm and 210 nm. Figure \ref{fig:sem} a)-d) show scanning electron microscope (SEM) images of the devices with close-ups of the grating coupler and the disk-waveguide coupling region. Figure \ref{fig:sem} e) shows a schematic cross-sectional view of the device, depicting the disk underetch, waveguide suspension, and the QW removal. More SEM close-ups of different components of the photonic circuit are shown in figure S10 in the additional microscope and SEM images section in the supporting information. A visualization of the top-view of a disk with a $90^\circ$ bent waveguide and its simulated $H_z$ field, showing the main device parameters, is depicted in figure S1 in the FDTD simulations section in the supporting information.

\begin{figure}
\includegraphics[width=1\linewidth]{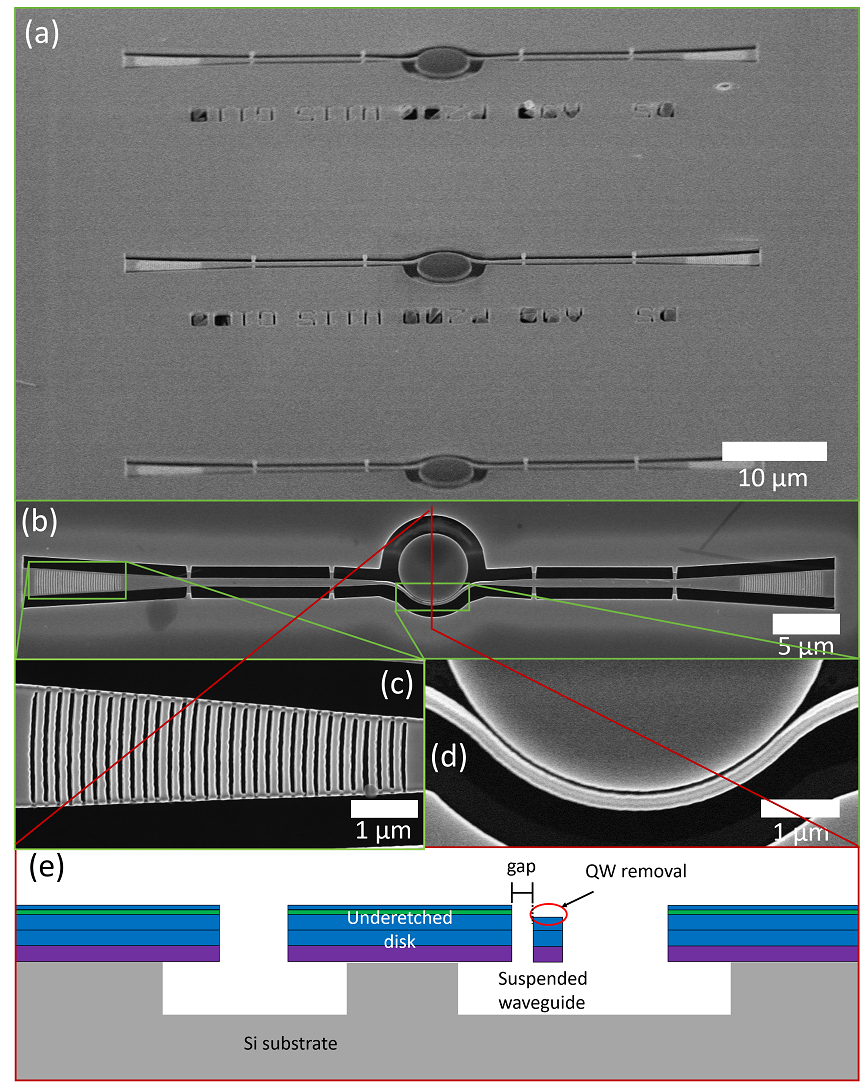}
  \caption{SEM image of III-nitride photonic circuit. a) View of several devices at an angle, b) top view of one device, c) close-up of a grating coupler, d) close-up of the disk and waveguide coupling region, e) sketch of a side-view of the fabricated device.}
  \label{fig:sem}
\end{figure}

\begin{figure}
\includegraphics[width=1\linewidth]{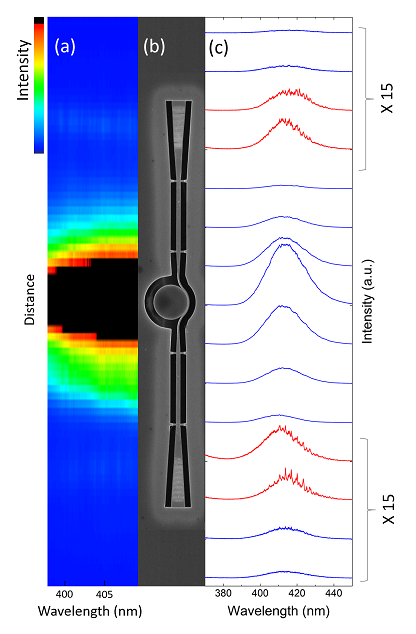}
  \caption{Visualization of the spatial distribution of the emission. a) 2D intensity map of the CCD, b) SEM image of the device matched to the CCD image along the vertical axis, c) spectra integrated over 5 pixels along the vertical axis corresponding to a) and b) in vertical position.}
  \label{fig:CCD}
\end{figure}

We performed micro-photoluminescence ($\mu$-PL) measurements on these devices using a low-power cw pump laser at 244 nm and a charged couple device (CCD) as the detector in a top-collection setup. Figure \ref{fig:CCD} a) shows a 2D map of the CCD, where the vertical axis is the distance along the device direction and the horizontal axis is the wavelength. In vertical direction 1 pixel on the CCD map corresponds to a distance of $0.9~\mu\text{m}$ on the sample. The vertical position on the CCD is matched with the SEM image next to it (figure \ref{fig:CCD} b)). The central, over-saturated emission comes from the QW luminescence scattered from the center of the microdisk. Faint emission, spatially separated from the high intensity central region can be observed from the grating couplers. Figure S8 in the additional micro-PL results section in the supporting information shows a more over-saturated version of figure \ref{fig:CCD} a) to highlight the emission from the gratings.  Figure \ref{fig:CCD} c) shows spectra integrated over 5 pixels along the vertical axis of the CCD, which corresponds to $4.5 ~\mu \text{m}$ on the sample. Clear whispering gallery mode (WGM) emission is observed from both gratings, while much higher intensity emission without any visible modes is observed from the center of the disk. The free carrier absorption in the n-doped part of the waveguide is only in the range of $50~\text{cm}^{-1}$ \cite{Kioupakis2010}. Taking the overlap of the mode with this region (28\%) into account, the estimated propagation losses due to doping are negligible.  A comparison between experimental and simulated spectra is shown in figure S2 in the FDTD simulations section in the supporting information. This result demonstrates light routing and extraction using an active blue emitter.

\begin{figure}
\includegraphics[width=1\linewidth]{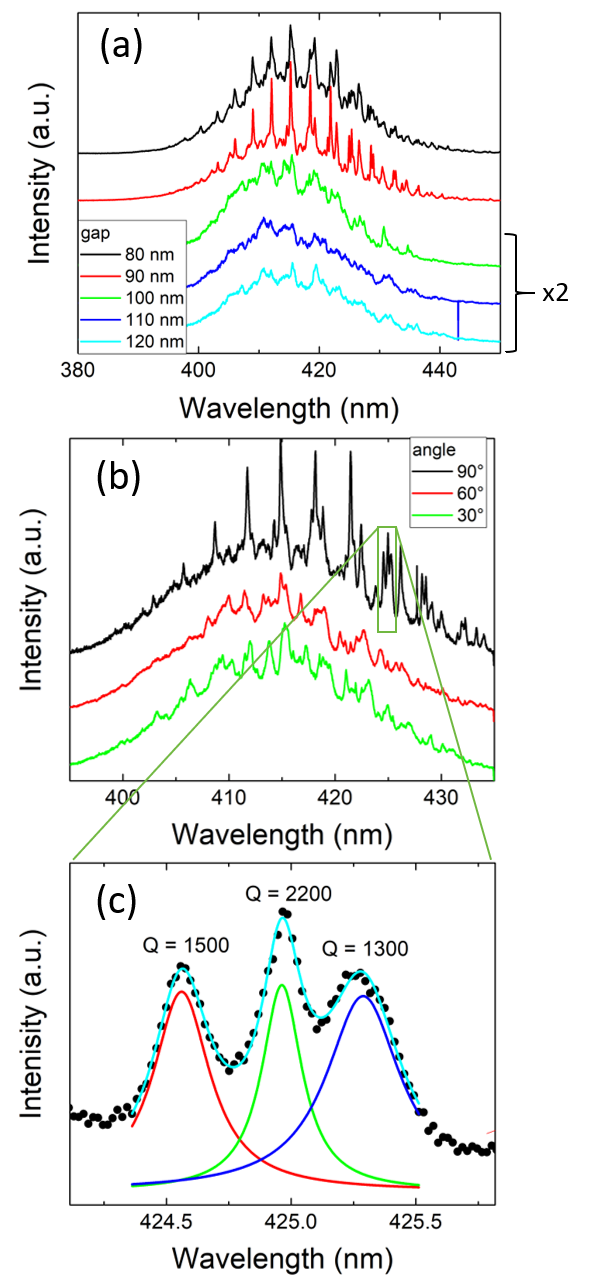}
  \caption{PL measurements integrated over one grating coupler on the CCD for devices with a disk diameter of $5~\mu \text{m}$, a waveguide width of 155 nm and a grating period of $210 ~\text{nm}$. a)  The waveguide bending angle is $90^\circ$ and the gap sizes vary between 80 nm and 120 nm. The spectra for $g = 100-120 ~\text{nm}$ have been multiplied by 2 for better visibility of the modes. b) The gap is 90 nm and the bending angle has been varied between  $90^\circ$ and  $30^\circ$. c) Close-up of three modes of the $90^\circ$ spectrum in b) showing Lorentzian fits and the corresponding Q factors.}
  \label{fig:PL}
\end{figure}

$\mu$-PL measurements for different gap sizes and bending angles are shown in figure \ref{fig:PL} for devices with $5~\mu\text{m}$ diameter disks, integrated spatially over one grating coupler on the CCD. Broad QW luminescence that couples to the waveguide and is outcoupled at the grating is observed in all spectra and decreases in intensity with increasing gap size. WGMs can also be seen in all spectra. Larger contrast WGM resonances are observed for small gaps (figure \ref{fig:PL} a)) and large angles (figure \ref{fig:PL} b)), giving loaded Q factors larger than 2000 in the blue, as can be seen in the close-up in figure \ref{fig:PL} c). We attribute the sharp resonances with high Q factor to the first-order radial modes. The azimuthal number for the first order radial mode at 422 nm is 84, according to our finite-difference time-domain (FDTD) simulations. Lower Q factor broader modes of higher radial order are also detected, indicating the presence of different families of modes. The here observed WGM contrast (defined as mode intensity divided by background intensity) of up to $2.7$ is the largest we have measured for such microdisks in a top-collection setup. In previous measurements, we had to use in-plane side-collection to detect the WGMs directly from the disk, since the modes radiate preferentially in the layer plane\cite{Mexis2011,Selles20162}. The loaded Q factors are very similar compared to the intrinsic Q factors of individual disks, which we have previously determined to be around 2500 for disks fabricated from the same wafer \cite{Selles20162}. Figure S6 in the additional micro-PL section in the supporting information shows measurements of  $3~\mu\text{m}$ diameter disks with $90^\circ$ bent waveguides depicting loaded Q factors in the range of 1200 to 1900.

The loaded Q factor for a disk and side-coupled waveguide is $1/Q_{loaded} = 1/Q_{int} + 1/Q_c$, where $Q_{int}$ is the intrinsic Q factor of the disk and $Q_c$ is the coupling Q factor. $Q_c$ depends on the mode overlap between the WGM and the waveguide mode, the interaction length and the phase mismatch. It is thus dependent on several parameters including the gap distance, the microdisk radius and the waveguide width. The spatial overlap of the first order radial mode with the waveguide mode is smaller than the one of higher-order radial modes. As shown by Soltani \cite{Soltani2009} for the case of silicon-based coupled resonators, the coupling strength (defined as $\frac{1}{Q_c}$) is usually weaker for the the first order radial modes thus leading to higher $Q_c$ factors as compared to higher-order radial modes. We mention the term "usually" as there might be some specific parameter combinations where this assumption does not hold because of phase mismatch issues. This feature has a strong impact on the mode visibility since the emission coupling to the waveguide depends on the values of $Q_{int}$ and $Q_c$ as for the case of transmission with microdisks and bus waveguides \cite{Yariv2000}.

As the gap distance increases,  $Q_c$ will increase exponentially and the fraction of the light transmitted through the waveguide drops very rapidly to zero. According to FDTD modeling (see figure S3 a) in the FDTD simulations section in the supporting information), the critical coupling distance is 50 nm for the first-order radial modes, i.e. the ones where  $Q_{int} =Q_c$ and thus $Q_{loaded} =1/2 Q_{int}$. For gap distances larger than 80 nm, we are in the under-coupled regime and the loaded Q factor should increase steadily towards the intrinsic Q factor. However, the  coupled intensity decreases. In the case of the 120 nm gap, the first-order radial modes are barely visible in FDTD transmission simulations (see figure S3 a) in the FDTD simulations section in the supporting information). Meanwhile, the higher-order radial modes have both a lower intrinsic quality factor and a lower coupling quality factor as compared to the first-order radial modes. The  amplitude of the higher-order radial modes coupled to the waveguide decreases as well with the gap distance but less sharply than the first-order radial modes. It explains why in figure \ref{fig:PL} a) the transmission spectra for gaps above 100 nm are dominated by the low Q factor broad modes and that the mode visibility of the first-order radial modes with high Q factor is quenched above 90 nm. Moreover, there is some scattered spontaneous emission from the disk collected above the waveguide and this background signal decreases the overall visibility of the modes. In future experiments, we will increase the distance between disk and grating to reduce the background signal.

\begin{figure}
\includegraphics[width=1\linewidth]{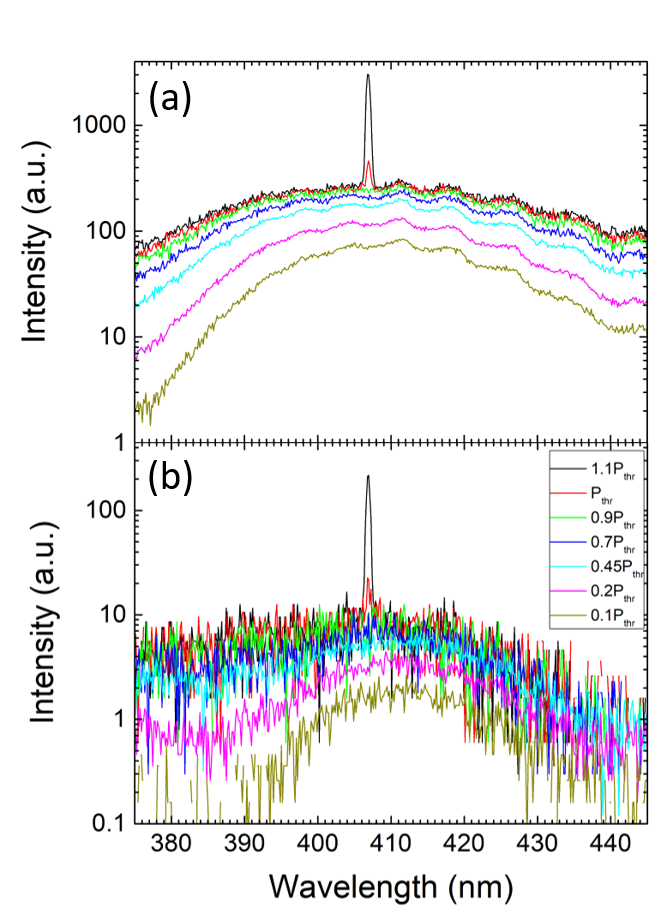}
  \caption{Logarithmic power-dependent pulsed lasing spectra of a device with a $3~\mu \text{m}$ diameter disk, a straight waveguide and a 120 nm gap measured a) above the disk and b) above the out-coupling grating. The threshold excitation power is $P_{thr}=15 ~\text{mJ/cm}^2$ per pulse. }
  \label{fig:lasing}
\end{figure}

Using a 266 nm laser under pulsed conditions (400 ps, 4 kHz), lasing was observed from devices with $3~\mu \text{m}$ diameter disks, straight waveguides and 100 to 120 nm gaps. Figure \ref{fig:lasing} shows power-dependent measurements both collected close to the disk (figure \ref{fig:lasing} a)) and from the out-coupling grating (figure \ref{fig:lasing} b)) for a device with a gap of 120 nm. The threshold was estimated to be $15 ~\text{mJ/cm}^2$ per pulse and lasing was observed up to our maximum available energy density of $16.5 ~\text{mJ/cm}^2$ per pulse. For individual microdisk lasers with a slightly larger diameter of $4~\mu \text{m}$ fabricated from the same wafer, we have previously observed thresholds of $3 ~\text{mJ/cm}^2$ per pulse \cite{Selles20162}. Figure S7 in the additional micro-PL section in the supporting information shows an S-shaped increase in lasing mode amplitude and linewidth narrowing for a $4~\mu\text{m}$ diameter disk without a waveguide fabricated on the same wafer. The here larger threshold is explained by differences in  thermal management and heat dissipation issues due to the lack of a silicon pedestal (see the microscope image in figure S9 a) in the additional microscope and SEM images section in the supporting information) \cite{Mexis2011} and fluctuations  in the processing with increased side-wall roughness and lower Q factors. The lasing mode and broad QW luminescence can be observed from both the disk and the grating. At the disk the emission has a higher intensity, while a factor two better contrast between the lasing mode and the QW luminescence is obtained at the grating. 

We do not observe lasing from any $5~\mu \text{m}$ diameter disks, which can be explained by an increase in threshold with increasing disk diameter and a decentralized pedestal near the disk edge causing additional losses. Furthermore, we also do not observe lasing from the devices with $3~\mu\text{m}$ diameter and $90^\circ$ bent waveguides due to losses in the pedestal, which is decentralized and near the disks edge in this case.

Based on FDTD simulations, we estimate the grating out-coupling efficiency to be 3-7\% (as shown in figure S5 in the FDTD  simulations section in the supporting information). The highest reported value in literature for a III-nitride grating coupler in the blue is 9\% \cite{Stegmaier2014}. The partial etching of the waveguide increases the coupling strength slightly, as shown by the FDTD simulations in figure S4 in the FDTD simulations section in the supporting information.

An interesting feature is that the lasing observed here is mono-mode, whereas we previously observed multi-mode lasing for single uncoupled microdisks \cite{Selles2016,Selles20162}. Mode selection is certainly enabled by the presence of the bus waveguide in close proximity with the microdisk, which might be the signature of a mechanism of mode selection involving the global structure (microdisk, waveguide and tethers). This will be the topic of future investigation. The measurements reported in figure \ref{fig:lasing}  along with the images shown in figure \ref{fig:sem} and figure \ref{fig:CCD} demonstrate the first integrated microlaser in the blue.

Under low-power cw excitation we did not observe any modes from devices with straight waveguides, as seen in figure \ref{fig:PL} for bent waveguides (not shown). This is explained by the weak coupling between the microdisk and the straight waveguide that prevents the observation of  modes. The bent geometry increases the interaction length and is known to increase the coupling when we are close to phase matching. This enhanced coupling is illustrated in figure \ref{fig:PL} b). The main advantage of the bent geometry is to allow a strong coupling, i.e. a low $Q_c$ value, for larger gap distances as compared to straight waveguides \cite{Hu2008}. The observation of the first-order radial modes with straight waveguides thus requires much smaller gap distances that are technologically more difficult to achieve with III-nitrides in the blue. It is furthermore challenging to etch III-nitrides with good verticality and smoothness in small openings. When the microdisk starts to lase, the amplitude becomes significantly stronger and the emission can be more easily observed after coupling to the waveguide and outcoupling through the grating.

\section{Conclusion}

We have demonstrated the first blue microlaser integrated into a photonic circuit using III-nitride on silicon. Large loaded Q factors greater than 2000 were observed at the outcoupling grating for devices with small gaps and large waveguide bending angles. Pulsed lasing was observed for devices with large gaps and straight waveguides with a threshold of $15 ~\text{mJ/cm}^2$ per pulse. Further work needs to be done on reducing the lasing threshold, increasing the coupling efficiency, and measuring / lowering transmission losses. The here presented results are an important step in demonstrating the viability of the III-nitride on silicon nanophotonic platform for real-world applications at short wavelength. A further challenge in making this photonic circuit platform on silicon viable for application is efficient electrical injection in III-nitride microdisks.

\begin{acknowledgments}
This work was supported by Agence Nationale de la Recherche under
MILAGAN convention (ANR-17-CE08-0043-02). This work was also partly
supported by the RENATECH network. We acknowledge support by a public
grant overseen by the French National Research Agency (ANR) as part
of the \textquotedblleft Investissements d\textquoteright Avenir\textquotedblright{}
program: Labex GANEX (Grant No. ANR-11-LABX-0014) and Labex NanoSaclay
(reference: ANR-10-LABX-0035). We thank Isabelle Robert-Philip for fruitful discussions.
\end{acknowledgments}

\bibliography{mybib}

\begin{thebibliography}{33}%
\makeatletter
\providecommand \@ifxundefined [1]{%
 \@ifx{#1\undefined}
}%
\providecommand \@ifnum [1]{%
 \ifnum #1\expandafter \@firstoftwo
 \else \expandafter \@secondoftwo
 \fi
}%
\providecommand \@ifx [1]{%
 \ifx #1\expandafter \@firstoftwo
 \else \expandafter \@secondoftwo
 \fi
}%
\providecommand \natexlab [1]{#1}%
\providecommand \enquote  [1]{``#1''}%
\providecommand \bibnamefont  [1]{#1}%
\providecommand \bibfnamefont [1]{#1}%
\providecommand \citenamefont [1]{#1}%
\providecommand \href@noop [0]{\@secondoftwo}%
\providecommand \href [0]{\begingroup \@sanitize@url \@href}%
\providecommand \@href[1]{\@@startlink{#1}\@@href}%
\providecommand \@@href[1]{\endgroup#1\@@endlink}%
\providecommand \@sanitize@url [0]{\catcode `\\12\catcode `\$12\catcode
  `\&12\catcode `\#12\catcode `\^12\catcode `\_12\catcode `\%12\relax}%
\providecommand \@@startlink[1]{}%
\providecommand \@@endlink[0]{}%
\providecommand \url  [0]{\begingroup\@sanitize@url \@url }%
\providecommand \@url [1]{\endgroup\@href {#1}{\urlprefix }}%
\providecommand \urlprefix  [0]{URL }%
\providecommand \Eprint [0]{\href }%
\providecommand \doibase [0]{http://dx.doi.org/}%
\providecommand \selectlanguage [0]{\@gobble}%
\providecommand \bibinfo  [0]{\@secondoftwo}%
\providecommand \bibfield  [0]{\@secondoftwo}%
\providecommand \translation [1]{[#1]}%
\providecommand \BibitemOpen [0]{}%
\providecommand \bibitemStop [0]{}%
\providecommand \bibitemNoStop [0]{.\EOS\space}%
\providecommand \EOS [0]{\spacefactor3000\relax}%
\providecommand \BibitemShut  [1]{\csname bibitem#1\endcsname}%
\let\auto@bib@innerbib\@empty
\bibitem [{\citenamefont {Xiong}\ \emph
  {et~al.}(2012{\natexlab{a}})\citenamefont {Xiong}, \citenamefont {Pernice},
  ,\ and\ \citenamefont {Tang}}]{Xiong2012}%
  \BibitemOpen
  \bibfield  {author} {\bibinfo {author} {\bibfnamefont {C.}~\bibnamefont
  {Xiong}}, \bibinfo {author} {\bibfnamefont {W.~H.~P.}\ \bibnamefont
  {Pernice}}, , \ and\ \bibinfo {author} {\bibfnamefont {H.~X.}\ \bibnamefont
  {Tang}},\ }\href {\doibase 10.1021/nl3011885} {\bibfield  {journal} {\bibinfo
   {journal} {Nano Lett.}\ }\textbf {\bibinfo {volume} {12}},\ \bibinfo {pages}
  {3562} (\bibinfo {year} {2012}{\natexlab{a}})}\BibitemShut {NoStop}%
\bibitem [{\citenamefont {Pernice}\ \emph {et~al.}(2012)\citenamefont
  {Pernice}, \citenamefont {Xiong}, \citenamefont {Schuck},\ and\ \citenamefont
  {Tang}}]{Pernice2012}%
  \BibitemOpen
  \bibfield  {author} {\bibinfo {author} {\bibfnamefont {W.~H.~P.}\
  \bibnamefont {Pernice}}, \bibinfo {author} {\bibfnamefont {C.}~\bibnamefont
  {Xiong}}, \bibinfo {author} {\bibfnamefont {C.}~\bibnamefont {Schuck}}, \
  and\ \bibinfo {author} {\bibfnamefont {H.~X.}\ \bibnamefont {Tang}},\ }\href
  {\doibase 10.1063/1.4722941} {\bibfield  {journal} {\bibinfo  {journal}
  {Appl. Phys. Lett.}\ }\textbf {\bibinfo {volume} {100}},\ \bibinfo {pages}
  {223501} (\bibinfo {year} {2012})}\BibitemShut {NoStop}%
\bibitem [{\citenamefont {N\'eel}\ \emph {et~al.}(2014)\citenamefont {N\'eel},
  \citenamefont {Roland}, \citenamefont {Checoury}, \citenamefont
  {El$\thinspace$Kurdi}, \citenamefont {Sauvage}, \citenamefont {Brimont},
  \citenamefont {Guillet}, \citenamefont {Gayral}, \citenamefont {Semond},\
  and\ \citenamefont {Boucaud}}]{Neel2014}%
  \BibitemOpen
  \bibfield  {author} {\bibinfo {author} {\bibfnamefont {D.}~\bibnamefont
  {N\'eel}}, \bibinfo {author} {\bibfnamefont {I.}~\bibnamefont {Roland}},
  \bibinfo {author} {\bibfnamefont {X.}~\bibnamefont {Checoury}}, \bibinfo
  {author} {\bibfnamefont {M.}~\bibnamefont {El$\thinspace$Kurdi}}, \bibinfo
  {author} {\bibfnamefont {S.}~\bibnamefont {Sauvage}}, \bibinfo {author}
  {\bibfnamefont {C.}~\bibnamefont {Brimont}}, \bibinfo {author} {\bibfnamefont
  {T.}~\bibnamefont {Guillet}}, \bibinfo {author} {\bibfnamefont
  {B.}~\bibnamefont {Gayral}}, \bibinfo {author} {\bibfnamefont
  {F.}~\bibnamefont {Semond}}, \ and\ \bibinfo {author} {\bibfnamefont
  {P.}~\bibnamefont {Boucaud}},\ }\href {\doibase 10.1088/2043-6262/5/2/023001}
  {\bibfield  {journal} {\bibinfo  {journal} {Adv. Nat. Sci.: Nanosci.
  Nanotechnol.}\ }\textbf {\bibinfo {volume} {5}},\ \bibinfo {pages} {023001}
  (\bibinfo {year} {2014})}\BibitemShut {NoStop}%
\bibitem [{\citenamefont {Stegmaier}\ \emph {et~al.}(2014)\citenamefont
  {Stegmaier}, \citenamefont {Ebert}, \citenamefont {Meckbach}, \citenamefont
  {Ilin}, \citenamefont {Siegel},\ and\ \citenamefont
  {Pernice}}]{Stegmaier2014}%
  \BibitemOpen
  \bibfield  {author} {\bibinfo {author} {\bibfnamefont {M.}~\bibnamefont
  {Stegmaier}}, \bibinfo {author} {\bibfnamefont {J.}~\bibnamefont {Ebert}},
  \bibinfo {author} {\bibfnamefont {J.~M.}\ \bibnamefont {Meckbach}}, \bibinfo
  {author} {\bibfnamefont {K.}~\bibnamefont {Ilin}}, \bibinfo {author}
  {\bibfnamefont {M.}~\bibnamefont {Siegel}}, \ and\ \bibinfo {author}
  {\bibfnamefont {W.~H.~P.}\ \bibnamefont {Pernice}},\ }\href {\doibase
  10.1063/1.4867529} {\bibfield  {journal} {\bibinfo  {journal} {Appl. Phys.
  Lett.}\ }\textbf {\bibinfo {volume} {104}},\ \bibinfo {pages} {091108}
  (\bibinfo {year} {2014})}\BibitemShut {NoStop}%
\bibitem [{\citenamefont {Roland}\ \emph {et~al.}(2016)\citenamefont {Roland},
  \citenamefont {Zeng}, \citenamefont {Checoury}, \citenamefont
  {El$\thinspace$Kurdi}, \citenamefont {Sauvage}, \citenamefont {Brimont},
  \citenamefont {Guillet}, \citenamefont {Gayral}, \citenamefont {Gromovyi},
  \citenamefont {Duboz}, \citenamefont {Semond}, \citenamefont {de~Micheli},\
  and\ \citenamefont {Boucaud}}]{Roland2016}%
  \BibitemOpen
  \bibfield  {author} {\bibinfo {author} {\bibfnamefont {I.}~\bibnamefont
  {Roland}}, \bibinfo {author} {\bibfnamefont {Y.}~\bibnamefont {Zeng}},
  \bibinfo {author} {\bibfnamefont {X.}~\bibnamefont {Checoury}}, \bibinfo
  {author} {\bibfnamefont {M.}~\bibnamefont {El$\thinspace$Kurdi}}, \bibinfo
  {author} {\bibfnamefont {S.}~\bibnamefont {Sauvage}}, \bibinfo {author}
  {\bibfnamefont {C.}~\bibnamefont {Brimont}}, \bibinfo {author} {\bibfnamefont
  {T.}~\bibnamefont {Guillet}}, \bibinfo {author} {\bibfnamefont
  {B.}~\bibnamefont {Gayral}}, \bibinfo {author} {\bibfnamefont
  {M.}~\bibnamefont {Gromovyi}}, \bibinfo {author} {\bibfnamefont {J.~Y.}\
  \bibnamefont {Duboz}}, \bibinfo {author} {\bibfnamefont {F.}~\bibnamefont
  {Semond}}, \bibinfo {author} {\bibfnamefont {M.~P.}\ \bibnamefont
  {de~Micheli}}, \ and\ \bibinfo {author} {\bibfnamefont {P.}~\bibnamefont
  {Boucaud}},\ }\href {\doibase 10.1364/OE.24.009602} {\bibfield  {journal}
  {\bibinfo  {journal} {Optics Express}\ }\textbf {\bibinfo {volume} {24}},\
  \bibinfo {pages} {9602} (\bibinfo {year} {2016})}\BibitemShut {NoStop}%
\bibitem [{\citenamefont {Smit}\ \emph {et~al.}(2014)\citenamefont {Smit},
  \citenamefont {Leijtens}, \citenamefont {Ambrosius}, \citenamefont {Bente},
  \citenamefont {van~der Tol}, \citenamefont {Smalbrugge}, \citenamefont
  {de~Vries}, \citenamefont {Geluk}, \citenamefont {Bolk}, \citenamefont {van
  Veldhoven}, \citenamefont {Augustin}, \citenamefont {Thijs}, \citenamefont
  {D’Agostino}, \citenamefont {Rabbani}, \citenamefont {Lawniczuk},
  \citenamefont {Stopinski}, \citenamefont {Tahvili}, \citenamefont {Corradi},
  \citenamefont {Kleijn}, \citenamefont {Dzibrou}, \citenamefont {Felicetti},
  \citenamefont {Bitincka}, \citenamefont {Moskalenko}, \citenamefont {Zhao},
  \citenamefont {Santos}, \citenamefont {Gilardi}, \citenamefont {Yao},
  \citenamefont {Williams}, \citenamefont {Stabile}, \citenamefont
  {Kuindersma}, \citenamefont {Pello}, \citenamefont {Bhat}, \citenamefont
  {Jiao}, \citenamefont {Heiss}, \citenamefont {Roelkens}, \citenamefont
  {Wale}, \citenamefont {Firth}, \citenamefont {Soares}, \citenamefont {Grote},
  \citenamefont {Schell}, \citenamefont {Debregeas}, \citenamefont {Achouche},
  \citenamefont {Gentner}, \citenamefont {Bakker}, \citenamefont {Korthorst},
  \citenamefont {Gallagher}, \citenamefont {Dabbs}, \citenamefont {Melloni},
  \citenamefont {Morichetti}, \citenamefont {Melati}, \citenamefont {Wonfor},
  \citenamefont {Penty}, \citenamefont {Broeke}, \citenamefont {Musk},\ and\
  \citenamefont {Robbins}}]{Smit2014}%
  \BibitemOpen
  \bibfield  {author} {\bibinfo {author} {\bibfnamefont {M.}~\bibnamefont
  {Smit}}, \bibinfo {author} {\bibfnamefont {X.}~\bibnamefont {Leijtens}},
  \bibinfo {author} {\bibfnamefont {H.}~\bibnamefont {Ambrosius}}, \bibinfo
  {author} {\bibfnamefont {E.}~\bibnamefont {Bente}}, \bibinfo {author}
  {\bibfnamefont {J.}~\bibnamefont {van~der Tol}}, \bibinfo {author}
  {\bibfnamefont {B.}~\bibnamefont {Smalbrugge}}, \bibinfo {author}
  {\bibfnamefont {T.}~\bibnamefont {de~Vries}}, \bibinfo {author}
  {\bibfnamefont {E.-J.}\ \bibnamefont {Geluk}}, \bibinfo {author}
  {\bibfnamefont {J.}~\bibnamefont {Bolk}}, \bibinfo {author} {\bibfnamefont
  {R.}~\bibnamefont {van Veldhoven}}, \bibinfo {author} {\bibfnamefont
  {L.}~\bibnamefont {Augustin}}, \bibinfo {author} {\bibfnamefont
  {P.}~\bibnamefont {Thijs}}, \bibinfo {author} {\bibfnamefont
  {D.}~\bibnamefont {D’Agostino}}, \bibinfo {author} {\bibfnamefont
  {H.}~\bibnamefont {Rabbani}}, \bibinfo {author} {\bibfnamefont
  {K.}~\bibnamefont {Lawniczuk}}, \bibinfo {author} {\bibfnamefont
  {S.}~\bibnamefont {Stopinski}}, \bibinfo {author} {\bibfnamefont
  {S.}~\bibnamefont {Tahvili}}, \bibinfo {author} {\bibfnamefont
  {A.}~\bibnamefont {Corradi}}, \bibinfo {author} {\bibfnamefont
  {E.}~\bibnamefont {Kleijn}}, \bibinfo {author} {\bibfnamefont
  {D.}~\bibnamefont {Dzibrou}}, \bibinfo {author} {\bibfnamefont
  {M.}~\bibnamefont {Felicetti}}, \bibinfo {author} {\bibfnamefont
  {E.}~\bibnamefont {Bitincka}}, \bibinfo {author} {\bibfnamefont
  {V.}~\bibnamefont {Moskalenko}}, \bibinfo {author} {\bibfnamefont
  {J.}~\bibnamefont {Zhao}}, \bibinfo {author} {\bibfnamefont {R.}~\bibnamefont
  {Santos}}, \bibinfo {author} {\bibfnamefont {G.}~\bibnamefont {Gilardi}},
  \bibinfo {author} {\bibfnamefont {W.}~\bibnamefont {Yao}}, \bibinfo {author}
  {\bibfnamefont {K.}~\bibnamefont {Williams}}, \bibinfo {author}
  {\bibfnamefont {P.}~\bibnamefont {Stabile}}, \bibinfo {author} {\bibfnamefont
  {P.}~\bibnamefont {Kuindersma}}, \bibinfo {author} {\bibfnamefont
  {J.}~\bibnamefont {Pello}}, \bibinfo {author} {\bibfnamefont
  {S.}~\bibnamefont {Bhat}}, \bibinfo {author} {\bibfnamefont {Y.}~\bibnamefont
  {Jiao}}, \bibinfo {author} {\bibfnamefont {D.}~\bibnamefont {Heiss}},
  \bibinfo {author} {\bibfnamefont {G.}~\bibnamefont {Roelkens}}, \bibinfo
  {author} {\bibfnamefont {M.}~\bibnamefont {Wale}}, \bibinfo {author}
  {\bibfnamefont {P.}~\bibnamefont {Firth}}, \bibinfo {author} {\bibfnamefont
  {F.}~\bibnamefont {Soares}}, \bibinfo {author} {\bibfnamefont
  {N.}~\bibnamefont {Grote}}, \bibinfo {author} {\bibfnamefont
  {M.}~\bibnamefont {Schell}}, \bibinfo {author} {\bibfnamefont
  {H.}~\bibnamefont {Debregeas}}, \bibinfo {author} {\bibfnamefont
  {M.}~\bibnamefont {Achouche}}, \bibinfo {author} {\bibfnamefont {J.-L.}\
  \bibnamefont {Gentner}}, \bibinfo {author} {\bibfnamefont {A.}~\bibnamefont
  {Bakker}}, \bibinfo {author} {\bibfnamefont {T.}~\bibnamefont {Korthorst}},
  \bibinfo {author} {\bibfnamefont {D.}~\bibnamefont {Gallagher}}, \bibinfo
  {author} {\bibfnamefont {A.}~\bibnamefont {Dabbs}}, \bibinfo {author}
  {\bibfnamefont {A.}~\bibnamefont {Melloni}}, \bibinfo {author} {\bibfnamefont
  {F.}~\bibnamefont {Morichetti}}, \bibinfo {author} {\bibfnamefont
  {D.}~\bibnamefont {Melati}}, \bibinfo {author} {\bibfnamefont
  {A.}~\bibnamefont {Wonfor}}, \bibinfo {author} {\bibfnamefont
  {R.}~\bibnamefont {Penty}}, \bibinfo {author} {\bibfnamefont
  {R.}~\bibnamefont {Broeke}}, \bibinfo {author} {\bibfnamefont
  {B.}~\bibnamefont {Musk}}, \ and\ \bibinfo {author} {\bibfnamefont
  {D.}~\bibnamefont {Robbins}},\ }\href {\doibase
  10.1088/0268-1242/29/8/083001} {\bibfield  {journal} {\bibinfo  {journal}
  {Semicond. Sci. Technol.}\ }\textbf {\bibinfo {volume} {29}},\ \bibinfo
  {pages} {083001} (\bibinfo {year} {2014})}\BibitemShut {NoStop}%
\bibitem [{\citenamefont {Guo}\ \emph {et~al.}(2017)\citenamefont {Guo},
  \citenamefont {ling Zou}, \citenamefont {Schuck}, \citenamefont {Jung},
  \citenamefont {Cheng},\ and\ \citenamefont {Tang}}]{Guo2017}%
  \BibitemOpen
  \bibfield  {author} {\bibinfo {author} {\bibfnamefont {X.}~\bibnamefont
  {Guo}}, \bibinfo {author} {\bibfnamefont {C.}~\bibnamefont {ling Zou}},
  \bibinfo {author} {\bibfnamefont {C.}~\bibnamefont {Schuck}}, \bibinfo
  {author} {\bibfnamefont {H.}~\bibnamefont {Jung}}, \bibinfo {author}
  {\bibfnamefont {R.}~\bibnamefont {Cheng}}, \ and\ \bibinfo {author}
  {\bibfnamefont {H.~X.}\ \bibnamefont {Tang}},\ }\href {\doibase
  10.1038/lsa.2016.249} {\bibfield  {journal} {\bibinfo  {journal} {Light:
  Science \& Applications}\ }\textbf {\bibinfo {volume} {6}},\ \bibinfo {pages}
  {e16249} (\bibinfo {year} {2017})}\BibitemShut {NoStop}%
\bibitem [{\citenamefont {Leonardis}\ \emph {et~al.}(2017)\citenamefont
  {Leonardis}, \citenamefont {Soref}, \citenamefont {Soltani},\ and\
  \citenamefont {Passaro}}]{Leonardis2017}%
  \BibitemOpen
  \bibfield  {author} {\bibinfo {author} {\bibfnamefont {F.~D.}\ \bibnamefont
  {Leonardis}}, \bibinfo {author} {\bibfnamefont {R.~A.}\ \bibnamefont
  {Soref}}, \bibinfo {author} {\bibfnamefont {M.}~\bibnamefont {Soltani}}, \
  and\ \bibinfo {author} {\bibfnamefont {V.~M.~N.}\ \bibnamefont {Passaro}},\
  }\href {\doibase 10.1038/s41598-017-11617-y} {\bibfield  {journal} {\bibinfo
  {journal} {Scientific Reports}\ }\textbf {\bibinfo {volume} {7}},\ \bibinfo
  {pages} {11387} (\bibinfo {year} {2017})}\BibitemShut {NoStop}%
\bibitem [{\citenamefont {Grosso}\ \emph {et~al.}(2017)\citenamefont {Grosso},
  \citenamefont {Lienhard}, \citenamefont {Moon}, \citenamefont {Scarabell},
  \citenamefont {Schroeder}, \citenamefont {Jeong}, \citenamefont {Lu},
  \citenamefont {Berhane}, \citenamefont {Wind}, \citenamefont {Aharanovich},\
  and\ \citenamefont {Englund}}]{Grosso2017}%
  \BibitemOpen
  \bibfield  {author} {\bibinfo {author} {\bibfnamefont {G.}~\bibnamefont
  {Grosso}}, \bibinfo {author} {\bibfnamefont {B.}~\bibnamefont {Lienhard}},
  \bibinfo {author} {\bibfnamefont {H.}~\bibnamefont {Moon}}, \bibinfo {author}
  {\bibfnamefont {D.}~\bibnamefont {Scarabell}}, \bibinfo {author}
  {\bibfnamefont {T.}~\bibnamefont {Schroeder}}, \bibinfo {author}
  {\bibfnamefont {K.-Y.}\ \bibnamefont {Jeong}}, \bibinfo {author}
  {\bibfnamefont {T.-J.}\ \bibnamefont {Lu}}, \bibinfo {author} {\bibfnamefont
  {A.~M.}\ \bibnamefont {Berhane}}, \bibinfo {author} {\bibfnamefont
  {S.}~\bibnamefont {Wind}}, \bibinfo {author} {\bibfnamefont {I.}~\bibnamefont
  {Aharanovich}}, \ and\ \bibinfo {author} {\bibfnamefont {D.}~\bibnamefont
  {Englund}},\ }\href {\doibase 10.1109/PHOSST.2017.8012671} {\bibfield
  {journal} {\bibinfo  {journal} {Photonics Society Summer Topical Meeting
  Series (SUM), IEEE}\ ,\ \bibinfo {pages} {103}} (\bibinfo {year}
  {2017})}\BibitemShut {NoStop}%
\bibitem [{\citenamefont {Berhane}\ \emph {et~al.}(2017)\citenamefont
  {Berhane}, \citenamefont {Jeong}, \citenamefont {Bodrog}, \citenamefont
  {Fiedler}, \citenamefont {Schr\"oder}, \citenamefont
  {Vico$\thinspace$Trivi{\~{n}}o}, \citenamefont {Palacios}, \citenamefont
  {Gali}, \citenamefont {Toth}, \citenamefont {Englund},\ and\ \citenamefont
  {Aharonovich}}]{Berhane2017}%
  \BibitemOpen
  \bibfield  {author} {\bibinfo {author} {\bibfnamefont {A.~M.}\ \bibnamefont
  {Berhane}}, \bibinfo {author} {\bibfnamefont {K.-Y.}\ \bibnamefont {Jeong}},
  \bibinfo {author} {\bibfnamefont {Z.}~\bibnamefont {Bodrog}}, \bibinfo
  {author} {\bibfnamefont {S.}~\bibnamefont {Fiedler}}, \bibinfo {author}
  {\bibfnamefont {T.}~\bibnamefont {Schr\"oder}}, \bibinfo {author}
  {\bibfnamefont {N.}~\bibnamefont {Vico$\thinspace$Trivi{\~{n}}o}}, \bibinfo
  {author} {\bibfnamefont {T.}~\bibnamefont {Palacios}}, \bibinfo {author}
  {\bibfnamefont {A.}~\bibnamefont {Gali}}, \bibinfo {author} {\bibfnamefont
  {M.}~\bibnamefont {Toth}}, \bibinfo {author} {\bibfnamefont {D.}~\bibnamefont
  {Englund}}, \ and\ \bibinfo {author} {\bibfnamefont {I.}~\bibnamefont
  {Aharonovich}},\ }\href {\doibase 10.1002/adma.201605092} {\bibfield
  {journal} {\bibinfo  {journal} {Adv. Mater.}\ }\textbf {\bibinfo {volume}
  {29}},\ \bibinfo {pages} {1605092} (\bibinfo {year} {2017})}\BibitemShut
  {NoStop}%
\bibitem [{\citenamefont {Jewett}\ \emph {et~al.}(2012)\citenamefont {Jewett},
  \citenamefont {Makowski}, \citenamefont {Andrews}, \citenamefont {Manfra},\
  and\ \citenamefont {Ivanisevic}}]{Jewett2012}%
  \BibitemOpen
  \bibfield  {author} {\bibinfo {author} {\bibfnamefont {S.~A.}\ \bibnamefont
  {Jewett}}, \bibinfo {author} {\bibfnamefont {M.~S.}\ \bibnamefont
  {Makowski}}, \bibinfo {author} {\bibfnamefont {B.}~\bibnamefont {Andrews}},
  \bibinfo {author} {\bibfnamefont {M.~J.}\ \bibnamefont {Manfra}}, \ and\
  \bibinfo {author} {\bibfnamefont {A.}~\bibnamefont {Ivanisevic}},\ }\href
  {\doibase 10.1016/j.actbio.2011.09.038} {\bibfield  {journal} {\bibinfo
  {journal} {Acta Biomaterialia}\ }\textbf {\bibinfo {volume} {8}},\ \bibinfo
  {pages} {728} (\bibinfo {year} {2012})}\BibitemShut {NoStop}%
\bibitem [{\citenamefont {Hofstetter}\ \emph {et~al.}(2012)\citenamefont
  {Hofstetter}, \citenamefont {Howgate}, \citenamefont {Schmid}, \citenamefont
  {Schoell}, \citenamefont {Sachsenhauser}, \citenamefont {Adig\'uzel},
  \citenamefont {Stutzmann}, \citenamefont {Sharp},\ and\ \citenamefont
  {Thalhammer}}]{Hofstetter2012}%
  \BibitemOpen
  \bibfield  {author} {\bibinfo {author} {\bibfnamefont {M.}~\bibnamefont
  {Hofstetter}}, \bibinfo {author} {\bibfnamefont {J.}~\bibnamefont {Howgate}},
  \bibinfo {author} {\bibfnamefont {M.}~\bibnamefont {Schmid}}, \bibinfo
  {author} {\bibfnamefont {S.}~\bibnamefont {Schoell}}, \bibinfo {author}
  {\bibfnamefont {M.}~\bibnamefont {Sachsenhauser}}, \bibinfo {author}
  {\bibfnamefont {D.}~\bibnamefont {Adig\'uzel}}, \bibinfo {author}
  {\bibfnamefont {M.}~\bibnamefont {Stutzmann}}, \bibinfo {author}
  {\bibfnamefont {I.~D.}\ \bibnamefont {Sharp}}, \ and\ \bibinfo {author}
  {\bibfnamefont {S.}~\bibnamefont {Thalhammer}},\ }\href {\doibase
  10.1016/j.bbrc.2012.06.142} {\bibfield  {journal} {\bibinfo  {journal}
  {Biochemical and Biophysical Research Communications}\ }\textbf {\bibinfo
  {volume} {424}},\ \bibinfo {pages} {348} (\bibinfo {year}
  {2012})}\BibitemShut {NoStop}%
\bibitem [{\citenamefont {Chi}\ \emph {et~al.}(2015)\citenamefont {Chi},
  \citenamefont {Hsieh}, \citenamefont {Lin}, \citenamefont {Chen},
  \citenamefont {Huang}, \citenamefont {He}, \citenamefont {Ooi}, \citenamefont
  {DenBaars}, \citenamefont {Nakamura}, \citenamefont {Kuo},\ and\
  \citenamefont {Lin}}]{Chi2015}%
  \BibitemOpen
  \bibfield  {author} {\bibinfo {author} {\bibfnamefont {Y.-C.}\ \bibnamefont
  {Chi}}, \bibinfo {author} {\bibfnamefont {D.-H.}\ \bibnamefont {Hsieh}},
  \bibinfo {author} {\bibfnamefont {C.-Y.}\ \bibnamefont {Lin}}, \bibinfo
  {author} {\bibfnamefont {H.-Y.}\ \bibnamefont {Chen}}, \bibinfo {author}
  {\bibfnamefont {C.-Y.}\ \bibnamefont {Huang}}, \bibinfo {author}
  {\bibfnamefont {J.-H.}\ \bibnamefont {He}}, \bibinfo {author} {\bibfnamefont
  {B.}~\bibnamefont {Ooi}}, \bibinfo {author} {\bibfnamefont {S.~P.}\
  \bibnamefont {DenBaars}}, \bibinfo {author} {\bibfnamefont {S.}~\bibnamefont
  {Nakamura}}, \bibinfo {author} {\bibfnamefont {H.-C.}\ \bibnamefont {Kuo}}, \
  and\ \bibinfo {author} {\bibfnamefont {G.-R.}\ \bibnamefont {Lin}},\ }\href
  {\doibase 10.1038/srep18690} {\bibfield  {journal} {\bibinfo  {journal}
  {Scientific Reports}\ }\textbf {\bibinfo {volume} {5}},\ \bibinfo {pages}
  {18690} (\bibinfo {year} {2015})}\BibitemShut {NoStop}%
\bibitem [{\citenamefont {Shi}\ \emph {et~al.}(2017)\citenamefont {Shi},
  \citenamefont {Gao}, \citenamefont {Yuan}, \citenamefont {Zhang},
  \citenamefont {Jiang}, \citenamefont {Zhang}, \citenamefont {Jiang},
  \citenamefont {Zhu},\ and\ \citenamefont {Wang}}]{Shi2017}%
  \BibitemOpen
  \bibfield  {author} {\bibinfo {author} {\bibfnamefont {Z.}~\bibnamefont
  {Shi}}, \bibinfo {author} {\bibfnamefont {X.}~\bibnamefont {Gao}}, \bibinfo
  {author} {\bibfnamefont {J.}~\bibnamefont {Yuan}}, \bibinfo {author}
  {\bibfnamefont {S.}~\bibnamefont {Zhang}}, \bibinfo {author} {\bibfnamefont
  {Y.}~\bibnamefont {Jiang}}, \bibinfo {author} {\bibfnamefont
  {F.}~\bibnamefont {Zhang}}, \bibinfo {author} {\bibfnamefont
  {Y.}~\bibnamefont {Jiang}}, \bibinfo {author} {\bibfnamefont
  {H.}~\bibnamefont {Zhu}}, \ and\ \bibinfo {author} {\bibfnamefont
  {Y.}~\bibnamefont {Wang}},\ }\href {\doibase 10.1063/1.5010892} {\bibfield
  {journal} {\bibinfo  {journal} {Appl. Phys. Lett.}\ }\textbf {\bibinfo
  {volume} {111}},\ \bibinfo {pages} {241104} (\bibinfo {year}
  {2017})}\BibitemShut {NoStop}%
\bibitem [{\citenamefont {Pernice}, \citenamefont {Xiong},\ and\ \citenamefont
  {Tang}(2012)}]{Pernice20122}%
  \BibitemOpen
  \bibfield  {author} {\bibinfo {author} {\bibfnamefont {W.~H.}\ \bibnamefont
  {Pernice}}, \bibinfo {author} {\bibfnamefont {C.}~\bibnamefont {Xiong}}, \
  and\ \bibinfo {author} {\bibfnamefont {H.~X.}\ \bibnamefont {Tang}},\ }\href
  {\doibase 10.1364/OE.20.012261} {\bibfield  {journal} {\bibinfo  {journal}
  {Optics Express}\ }\textbf {\bibinfo {volume} {20}},\ \bibinfo {pages}
  {12261} (\bibinfo {year} {2012})}\BibitemShut {NoStop}%
\bibitem [{\citenamefont {Xiong}\ \emph
  {et~al.}(2012{\natexlab{b}})\citenamefont {Xiong}, \citenamefont {Pernice},
  \citenamefont {Sun}, \citenamefont {Schuck}, \citenamefont {Fong},\ and\
  \citenamefont {Tang}}]{Xiong20122}%
  \BibitemOpen
  \bibfield  {author} {\bibinfo {author} {\bibfnamefont {C.}~\bibnamefont
  {Xiong}}, \bibinfo {author} {\bibfnamefont {W.~H.~P.}\ \bibnamefont
  {Pernice}}, \bibinfo {author} {\bibfnamefont {X.}~\bibnamefont {Sun}},
  \bibinfo {author} {\bibfnamefont {C.}~\bibnamefont {Schuck}}, \bibinfo
  {author} {\bibfnamefont {K.~Y.}\ \bibnamefont {Fong}}, \ and\ \bibinfo
  {author} {\bibfnamefont {H.~X.}\ \bibnamefont {Tang}},\ }\href {\doibase
  10.1088/1367-2630/14/9/095014} {\bibfield  {journal} {\bibinfo  {journal}
  {New Journal of Physics}\ }\textbf {\bibinfo {volume} {14}},\ \bibinfo
  {pages} {095014} (\bibinfo {year} {2012}{\natexlab{b}})}\BibitemShut
  {NoStop}%
\bibitem [{\citenamefont {Tamboli}\ \emph {et~al.}(2007)\citenamefont
  {Tamboli}, \citenamefont {Haberer}, \citenamefont {Sharma}, \citenamefont
  {Lee}, \citenamefont {Nakamura},\ and\ \citenamefont {Hu}}]{Tamboli2007}%
  \BibitemOpen
  \bibfield  {author} {\bibinfo {author} {\bibfnamefont {A.~C.}\ \bibnamefont
  {Tamboli}}, \bibinfo {author} {\bibfnamefont {E.~D.}\ \bibnamefont
  {Haberer}}, \bibinfo {author} {\bibfnamefont {R.}~\bibnamefont {Sharma}},
  \bibinfo {author} {\bibfnamefont {K.~H.}\ \bibnamefont {Lee}}, \bibinfo
  {author} {\bibfnamefont {S.}~\bibnamefont {Nakamura}}, \ and\ \bibinfo
  {author} {\bibfnamefont {E.~L.}\ \bibnamefont {Hu}},\ }\href {\doibase
  10.1038/nphoton.2006.52} {\bibfield  {journal} {\bibinfo  {journal} {Nature
  Photonics}\ }\textbf {\bibinfo {volume} {1}},\ \bibinfo {pages} {61}
  (\bibinfo {year} {2007})}\BibitemShut {NoStop}%
\bibitem [{\citenamefont {Simeonov}\ \emph {et~al.}(2007)\citenamefont
  {Simeonov}, \citenamefont {Feltin}, \citenamefont {B\"uhlmann}, \citenamefont
  {Zhu}, \citenamefont {Castiglia}, \citenamefont {Mosca}, \citenamefont
  {Carlin}, \citenamefont {Butt\'e},\ and\ \citenamefont
  {Grandjean}}]{Simeonov2007}%
  \BibitemOpen
  \bibfield  {author} {\bibinfo {author} {\bibfnamefont {D.}~\bibnamefont
  {Simeonov}}, \bibinfo {author} {\bibfnamefont {E.}~\bibnamefont {Feltin}},
  \bibinfo {author} {\bibfnamefont {H.-J.}\ \bibnamefont {B\"uhlmann}},
  \bibinfo {author} {\bibfnamefont {T.}~\bibnamefont {Zhu}}, \bibinfo {author}
  {\bibfnamefont {A.}~\bibnamefont {Castiglia}}, \bibinfo {author}
  {\bibfnamefont {M.}~\bibnamefont {Mosca}}, \bibinfo {author} {\bibfnamefont
  {J.-F.}\ \bibnamefont {Carlin}}, \bibinfo {author} {\bibfnamefont
  {R.}~\bibnamefont {Butt\'e}}, \ and\ \bibinfo {author} {\bibfnamefont
  {N.}~\bibnamefont {Grandjean}},\ }\href {\doibase 10.1063/1.2460234}
  {\bibfield  {journal} {\bibinfo  {journal} {Appl. Phys. Lett.}\ }\textbf
  {\bibinfo {volume} {90}},\ \bibinfo {pages} {061106} (\bibinfo {year}
  {2007})}\BibitemShut {NoStop}%
\bibitem [{\citenamefont {Simeonov}\ \emph {et~al.}(2008)\citenamefont
  {Simeonov}, \citenamefont {Feltin}, \citenamefont {Altoukhov}, \citenamefont
  {Castiglia}, \citenamefont {Carlin}, \citenamefont {Butt\'e},\ and\
  \citenamefont {Grandjean}}]{Simeonov2008}%
  \BibitemOpen
  \bibfield  {author} {\bibinfo {author} {\bibfnamefont {D.}~\bibnamefont
  {Simeonov}}, \bibinfo {author} {\bibfnamefont {E.}~\bibnamefont {Feltin}},
  \bibinfo {author} {\bibfnamefont {A.}~\bibnamefont {Altoukhov}}, \bibinfo
  {author} {\bibfnamefont {A.}~\bibnamefont {Castiglia}}, \bibinfo {author}
  {\bibfnamefont {J.-F.}\ \bibnamefont {Carlin}}, \bibinfo {author}
  {\bibfnamefont {R.}~\bibnamefont {Butt\'e}}, \ and\ \bibinfo {author}
  {\bibfnamefont {N.}~\bibnamefont {Grandjean}},\ }\href {\doibase
  10.1063/1.2917452} {\bibfield  {journal} {\bibinfo  {journal} {Appl. Phys.
  Lett.}\ }\textbf {\bibinfo {volume} {92}},\ \bibinfo {pages} {171102}
  (\bibinfo {year} {2008})}\BibitemShut {NoStop}%
\bibitem [{\citenamefont {Mexis}\ \emph {et~al.}(2011)\citenamefont {Mexis},
  \citenamefont {Sergent}, \citenamefont {Guillet}, \citenamefont {Brimont},
  \citenamefont {Bretagnon}, \citenamefont {B.~Gil~a}, \citenamefont {Leroux},
  \citenamefont {Néel}, \citenamefont {David}, \citenamefont {Chécoury},\
  and\ \citenamefont {Boucaud}}]{Mexis2011}%
  \BibitemOpen
  \bibfield  {author} {\bibinfo {author} {\bibfnamefont {M.}~\bibnamefont
  {Mexis}}, \bibinfo {author} {\bibfnamefont {S.}~\bibnamefont {Sergent}},
  \bibinfo {author} {\bibfnamefont {T.}~\bibnamefont {Guillet}}, \bibinfo
  {author} {\bibfnamefont {C.}~\bibnamefont {Brimont}}, \bibinfo {author}
  {\bibfnamefont {T.}~\bibnamefont {Bretagnon}}, \bibinfo {author}
  {\bibfnamefont {d.~F.~S.}\ \bibnamefont {B.~Gil~a}}, \bibinfo {author}
  {\bibfnamefont {M.}~\bibnamefont {Leroux}}, \bibinfo {author} {\bibfnamefont
  {D.}~\bibnamefont {Néel}}, \bibinfo {author} {\bibfnamefont
  {S.}~\bibnamefont {David}}, \bibinfo {author} {\bibfnamefont
  {X.}~\bibnamefont {Chécoury}}, \ and\ \bibinfo {author} {\bibfnamefont
  {P.}~\bibnamefont {Boucaud}},\ }\href {\doibase 10.1364/OL.36.002203}
  {\bibfield  {journal} {\bibinfo  {journal} {Optics Letters}\ }\textbf
  {\bibinfo {volume} {36}},\ \bibinfo {pages} {2203} (\bibinfo {year}
  {2011})}\BibitemShut {NoStop}%
\bibitem [{\citenamefont {Aharonovich}\ \emph {et~al.}(2013)\citenamefont
  {Aharonovich}, \citenamefont {Woolf}, \citenamefont {Russell}, \citenamefont
  {Zhu}, \citenamefont {Niu}, \citenamefont {Kappers}, \citenamefont {Oliver},\
  and\ \citenamefont {Hu}}]{Aharonovich2013}%
  \BibitemOpen
  \bibfield  {author} {\bibinfo {author} {\bibfnamefont {I.}~\bibnamefont
  {Aharonovich}}, \bibinfo {author} {\bibfnamefont {A.}~\bibnamefont {Woolf}},
  \bibinfo {author} {\bibfnamefont {K.~J.}\ \bibnamefont {Russell}}, \bibinfo
  {author} {\bibfnamefont {T.}~\bibnamefont {Zhu}}, \bibinfo {author}
  {\bibfnamefont {N.}~\bibnamefont {Niu}}, \bibinfo {author} {\bibfnamefont
  {M.~J.}\ \bibnamefont {Kappers}}, \bibinfo {author} {\bibfnamefont {R.~A.}\
  \bibnamefont {Oliver}}, \ and\ \bibinfo {author} {\bibfnamefont {E.~L.}\
  \bibnamefont {Hu}},\ }\href {\doibase 10.1063/1.4813471} {\bibfield
  {journal} {\bibinfo  {journal} {Appl. Phys. Lett.}\ }\textbf {\bibinfo
  {volume} {103}},\ \bibinfo {pages} {021112} (\bibinfo {year}
  {2013})}\BibitemShut {NoStop}%
\bibitem [{\citenamefont {Athanasiou}\ \emph {et~al.}(2014)\citenamefont
  {Athanasiou}, \citenamefont {Smith}, \citenamefont {Liu},\ and\ \citenamefont
  {Wang}}]{Athanasiou2014}%
  \BibitemOpen
  \bibfield  {author} {\bibinfo {author} {\bibfnamefont {M.}~\bibnamefont
  {Athanasiou}}, \bibinfo {author} {\bibfnamefont {R.}~\bibnamefont {Smith}},
  \bibinfo {author} {\bibfnamefont {B.}~\bibnamefont {Liu}}, \ and\ \bibinfo
  {author} {\bibfnamefont {T.}~\bibnamefont {Wang}},\ }\href {\doibase
  10.1038/srep07250} {\bibfield  {journal} {\bibinfo  {journal} {Scientific
  Reports}\ }\textbf {\bibinfo {volume} {4}},\ \bibinfo {pages} {7250}
  (\bibinfo {year} {2014})}\BibitemShut {NoStop}%
\bibitem [{\citenamefont {Zhang}\ \emph {et~al.}(2014)\citenamefont {Zhang},
  \citenamefont {Zhang}, \citenamefont {Li}, \citenamefont {Cheung},
  \citenamefont {Fengand},\ and\ \citenamefont {Choi}}]{Zhang2015}%
  \BibitemOpen
  \bibfield  {author} {\bibinfo {author} {\bibfnamefont {Y.}~\bibnamefont
  {Zhang}}, \bibinfo {author} {\bibfnamefont {X.}~\bibnamefont {Zhang}},
  \bibinfo {author} {\bibfnamefont {K.~H.}\ \bibnamefont {Li}}, \bibinfo
  {author} {\bibfnamefont {Y.~F.}\ \bibnamefont {Cheung}}, \bibinfo {author}
  {\bibfnamefont {C.}~\bibnamefont {Fengand}}, \ and\ \bibinfo {author}
  {\bibfnamefont {H.~W.}\ \bibnamefont {Choi}},\ }\href {\doibase
  10.1002/pssa.201431745} {\bibfield  {journal} {\bibinfo  {journal} {Phys.
  Status Solidi A}\ }\textbf {\bibinfo {volume} {210}},\ \bibinfo {pages} {960}
  (\bibinfo {year} {2014})}\BibitemShut {NoStop}%
\bibitem [{\citenamefont {Sell\'es}\ \emph
  {et~al.}(2016{\natexlab{a}})\citenamefont {Sell\'es}, \citenamefont
  {Brimont}, \citenamefont {Cassabois}, \citenamefont {Valvin}, \citenamefont
  {Guillet}, \citenamefont {Roland}, \citenamefont {Zeng}, \citenamefont
  {Checoury}, \citenamefont {Boucaud}, \citenamefont {Mexis}, \citenamefont
  {Semond},\ and\ \citenamefont {Gayral}}]{Selles2016}%
  \BibitemOpen
  \bibfield  {author} {\bibinfo {author} {\bibfnamefont {J.}~\bibnamefont
  {Sell\'es}}, \bibinfo {author} {\bibfnamefont {C.}~\bibnamefont {Brimont}},
  \bibinfo {author} {\bibfnamefont {G.}~\bibnamefont {Cassabois}}, \bibinfo
  {author} {\bibfnamefont {P.}~\bibnamefont {Valvin}}, \bibinfo {author}
  {\bibfnamefont {T.}~\bibnamefont {Guillet}}, \bibinfo {author} {\bibfnamefont
  {I.}~\bibnamefont {Roland}}, \bibinfo {author} {\bibfnamefont
  {Y.}~\bibnamefont {Zeng}}, \bibinfo {author} {\bibfnamefont {X.}~\bibnamefont
  {Checoury}}, \bibinfo {author} {\bibfnamefont {P.}~\bibnamefont {Boucaud}},
  \bibinfo {author} {\bibfnamefont {M.}~\bibnamefont {Mexis}}, \bibinfo
  {author} {\bibfnamefont {F.}~\bibnamefont {Semond}}, \ and\ \bibinfo {author}
  {\bibfnamefont {B.}~\bibnamefont {Gayral}},\ }\href {\doibase
  10.1038/srep21650} {\bibfield  {journal} {\bibinfo  {journal} {Scientific
  Reports}\ }\textbf {\bibinfo {volume} {6}},\ \bibinfo {pages} {21650}
  (\bibinfo {year} {2016}{\natexlab{a}})}\BibitemShut {NoStop}%
\bibitem [{\citenamefont {Rousseau}\ \emph {et~al.}(2018)\citenamefont
  {Rousseau}, \citenamefont {Callsen}, \citenamefont {Jacopin}, \citenamefont
  {Carlin}, \citenamefont {Butt\'e},\ and\ \citenamefont
  {Grandjean}}]{Rousseau2018}%
  \BibitemOpen
  \bibfield  {author} {\bibinfo {author} {\bibfnamefont {I.}~\bibnamefont
  {Rousseau}}, \bibinfo {author} {\bibfnamefont {G.}~\bibnamefont {Callsen}},
  \bibinfo {author} {\bibfnamefont {G.}~\bibnamefont {Jacopin}}, \bibinfo
  {author} {\bibfnamefont {J.-F.}\ \bibnamefont {Carlin}}, \bibinfo {author}
  {\bibfnamefont {R.}~\bibnamefont {Butt\'e}}, \ and\ \bibinfo {author}
  {\bibfnamefont {N.}~\bibnamefont {Grandjean}},\ }\href {\doibase
  10.1063/1.5022150} {\bibfield  {journal} {\bibinfo  {journal} {J. Appl.
  Phys.}\ }\textbf {\bibinfo {volume} {123}},\ \bibinfo {pages} {113103}
  (\bibinfo {year} {2018})}\BibitemShut {NoStop}%
\bibitem [{\citenamefont {Koseki}\ \emph {et~al.}(2009)\citenamefont {Koseki},
  \citenamefont {Zhang}, \citenamefont {Greve},\ and\ \citenamefont
  {Yamamoto}}]{Koseki2009}%
  \BibitemOpen
  \bibfield  {author} {\bibinfo {author} {\bibfnamefont {S.}~\bibnamefont
  {Koseki}}, \bibinfo {author} {\bibfnamefont {B.}~\bibnamefont {Zhang}},
  \bibinfo {author} {\bibfnamefont {K.~D.}\ \bibnamefont {Greve}}, \ and\
  \bibinfo {author} {\bibfnamefont {Y.}~\bibnamefont {Yamamoto}},\ }\href
  {\doibase 10.1063/1.3078522} {\bibfield  {journal} {\bibinfo  {journal}
  {Appl. Phys. Lett.}\ }\textbf {\bibinfo {volume} {94}},\ \bibinfo {pages}
  {051110} (\bibinfo {year} {2009})}\BibitemShut {NoStop}%
\bibitem [{\citenamefont {Schmidt}\ \emph {et~al.}(2017)\citenamefont
  {Schmidt}, \citenamefont {Rieger}, \citenamefont {Trellenkamp}, \citenamefont
  {Gr\"utzmacher},\ and\ \citenamefont {Pawlis}}]{Schmidt2017}%
  \BibitemOpen
  \bibfield  {author} {\bibinfo {author} {\bibfnamefont {G.}~\bibnamefont
  {Schmidt}}, \bibinfo {author} {\bibfnamefont {T.}~\bibnamefont {Rieger}},
  \bibinfo {author} {\bibfnamefont {S.}~\bibnamefont {Trellenkamp}}, \bibinfo
  {author} {\bibfnamefont {D.}~\bibnamefont {Gr\"utzmacher}}, \ and\ \bibinfo
  {author} {\bibfnamefont {A.}~\bibnamefont {Pawlis}},\ }\href {\doibase
  10.1088/1361-6641/aa69f5} {\bibfield  {journal} {\bibinfo  {journal}
  {Semicond. Sci. Technol.}\ }\textbf {\bibinfo {volume} {32}},\ \bibinfo
  {pages} {075015} (\bibinfo {year} {2017})}\BibitemShut {NoStop}%
\bibitem [{\citenamefont {Matsuo}\ \emph {et~al.}(2010)\citenamefont {Matsuo},
  \citenamefont {Shinya}, \citenamefont {Kakitsuka}, \citenamefont {Nozaki},
  \citenamefont {Segawa}, \citenamefont {Sato}, \citenamefont {Kawaguchi},\
  and\ \citenamefont {Notomi}}]{Matsuo2010}%
  \BibitemOpen
  \bibfield  {author} {\bibinfo {author} {\bibfnamefont {S.}~\bibnamefont
  {Matsuo}}, \bibinfo {author} {\bibfnamefont {A.}~\bibnamefont {Shinya}},
  \bibinfo {author} {\bibfnamefont {T.}~\bibnamefont {Kakitsuka}}, \bibinfo
  {author} {\bibfnamefont {K.}~\bibnamefont {Nozaki}}, \bibinfo {author}
  {\bibfnamefont {T.}~\bibnamefont {Segawa}}, \bibinfo {author} {\bibfnamefont
  {T.}~\bibnamefont {Sato}}, \bibinfo {author} {\bibfnamefont {Y.}~\bibnamefont
  {Kawaguchi}}, \ and\ \bibinfo {author} {\bibfnamefont {M.}~\bibnamefont
  {Notomi}},\ }\href {\doibase 10.1038/nphoton.2010.177} {\bibfield  {journal}
  {\bibinfo  {journal} {Nature Photonics}\ }\textbf {\bibinfo {volume} {4}},\
  \bibinfo {pages} {648} (\bibinfo {year} {2010})}\BibitemShut {NoStop}%
\bibitem [{\citenamefont {Kioupakis}\ \emph {et~al.}(2010)\citenamefont
  {Kioupakis}, \citenamefont {Rinke}, \citenamefont {Schleife}, \citenamefont
  {Bechstedt},\ and\ \citenamefont
  {Van$\thinspace$de$\thinspace$Walle}}]{Kioupakis2010}%
  \BibitemOpen
  \bibfield  {author} {\bibinfo {author} {\bibfnamefont {E.}~\bibnamefont
  {Kioupakis}}, \bibinfo {author} {\bibfnamefont {P.}~\bibnamefont {Rinke}},
  \bibinfo {author} {\bibfnamefont {A.}~\bibnamefont {Schleife}}, \bibinfo
  {author} {\bibfnamefont {F.}~\bibnamefont {Bechstedt}}, \ and\ \bibinfo
  {author} {\bibfnamefont {C.~G.}\ \bibnamefont
  {Van$\thinspace$de$\thinspace$Walle}},\ }\href {\doibase
  10.1103/PhysRevB.81.241201} {\bibfield  {journal} {\bibinfo  {journal} {Phys.
  Rev. B}\ }\textbf {\bibinfo {volume} {81}},\ \bibinfo {pages} {241201(R)}
  (\bibinfo {year} {2010})}\BibitemShut {NoStop}%
\bibitem [{\citenamefont {Sell\'es}\ \emph
  {et~al.}(2016{\natexlab{b}})\citenamefont {Sell\'es}, \citenamefont {Crepel},
  \citenamefont {Roland}, \citenamefont {Kurdi}, \citenamefont {Checoury},
  \citenamefont {Boucaud}, \citenamefont {Mexis}, \citenamefont {Leroux},
  \citenamefont {Damilano}, \citenamefont {Rennesson}, \citenamefont {Semond},
  \citenamefont {Gayral}, \citenamefont {Brimont},\ and\ \citenamefont
  {Guillet}}]{Selles20162}%
  \BibitemOpen
  \bibfield  {author} {\bibinfo {author} {\bibfnamefont {J.}~\bibnamefont
  {Sell\'es}}, \bibinfo {author} {\bibfnamefont {V.}~\bibnamefont {Crepel}},
  \bibinfo {author} {\bibfnamefont {I.}~\bibnamefont {Roland}}, \bibinfo
  {author} {\bibfnamefont {M.~E.}\ \bibnamefont {Kurdi}}, \bibinfo {author}
  {\bibfnamefont {X.}~\bibnamefont {Checoury}}, \bibinfo {author}
  {\bibfnamefont {P.}~\bibnamefont {Boucaud}}, \bibinfo {author} {\bibfnamefont
  {M.}~\bibnamefont {Mexis}}, \bibinfo {author} {\bibfnamefont
  {M.}~\bibnamefont {Leroux}}, \bibinfo {author} {\bibfnamefont
  {B.}~\bibnamefont {Damilano}}, \bibinfo {author} {\bibfnamefont
  {S.}~\bibnamefont {Rennesson}}, \bibinfo {author} {\bibfnamefont
  {F.}~\bibnamefont {Semond}}, \bibinfo {author} {\bibfnamefont
  {B.}~\bibnamefont {Gayral}}, \bibinfo {author} {\bibfnamefont
  {C.}~\bibnamefont {Brimont}}, \ and\ \bibinfo {author} {\bibfnamefont
  {T.}~\bibnamefont {Guillet}},\ }\href {\doibase 10.1063/1.4971357} {\bibfield
   {journal} {\bibinfo  {journal} {Appl. Phys. Lett.}\ }\textbf {\bibinfo
  {volume} {109}},\ \bibinfo {pages} {231101} (\bibinfo {year}
  {2016}{\natexlab{b}})}\BibitemShut {NoStop}%
\bibitem [{\citenamefont {Soltani}(2009)}]{Soltani2009}%
  \BibitemOpen
  \bibfield  {author} {\bibinfo {author} {\bibfnamefont {M.}~\bibnamefont
  {Soltani}},\ }\emph {\bibinfo {title} {Novel Integrated Silicon Nanophotonic
  Structures using Ultra-high Q Resonators}},\ \href@noop {} {Ph.D. thesis},\
  \bibinfo  {school} {Georgia Institute of Technology} (\bibinfo {year}
  {2009})\BibitemShut {NoStop}%
\bibitem [{\citenamefont {Yariv}(2000)}]{Yariv2000}%
  \BibitemOpen
  \bibfield  {author} {\bibinfo {author} {\bibfnamefont {A.}~\bibnamefont
  {Yariv}},\ }\href {\doibase 10.1049/el:20000340} {\bibfield  {journal}
  {\bibinfo  {journal} {Electronics Letters}\ }\textbf {\bibinfo {volume}
  {36}},\ \bibinfo {pages} {321} (\bibinfo {year} {2000})}\BibitemShut
  {NoStop}%
\bibitem [{\citenamefont {Hu}\ \emph {et~al.}(2008)\citenamefont {Hu},
  \citenamefont {Carlie}, \citenamefont {Feng}, \citenamefont {Petit},
  \citenamefont {Agarwal}, \citenamefont {Richardson},\ and\ \citenamefont
  {Kimerling}}]{Hu2008}%
  \BibitemOpen
  \bibfield  {author} {\bibinfo {author} {\bibfnamefont {J.}~\bibnamefont
  {Hu}}, \bibinfo {author} {\bibfnamefont {N.}~\bibnamefont {Carlie}}, \bibinfo
  {author} {\bibfnamefont {N.-N.}\ \bibnamefont {Feng}}, \bibinfo {author}
  {\bibfnamefont {L.}~\bibnamefont {Petit}}, \bibinfo {author} {\bibfnamefont
  {A.}~\bibnamefont {Agarwal}}, \bibinfo {author} {\bibfnamefont
  {K.}~\bibnamefont {Richardson}}, \ and\ \bibinfo {author} {\bibfnamefont
  {L.}~\bibnamefont {Kimerling}},\ }\href {\doibase 10.1364/OL.33.002500}
  {\bibfield  {journal} {\bibinfo  {journal} {Opt. Lett.}\ }\textbf {\bibinfo
  {volume} {33}},\ \bibinfo {pages} {2500} (\bibinfo {year}
  {2008})}\BibitemShut {NoStop}%
\end{thebibliography}%


\newpage

\onecolumngrid
\begin{center}
  \textbf{\large Supporting Information for \\
  Blue microlasers integrated on a photonic platform on silicon}\\[.2cm]
  Farsane Tabataba-Vakili,$^{1,2}$ Laetitia Doyennette,$^{3}$ Christelle Brimont,$^3$ Thierry Guillet,$^3$ St\'{e}phanie Rennesson,$^4$ Eric Frayssinet,$^4$ Benjamin Damilano,$^4$ Jean-Yves Duboz,$^4$ Fabrice Semond,$^4$ Iannis Roland,$^{1,a)}$ Moustafa El Kurdi,$^1$ Xavier Checoury,$^1$ S\'ebastien Sauvage,$^1$ Bruno Gayral$^2$ and Philippe Boucaud$^{1,4,b)}$   \\[.1cm]
  {\itshape ${}^1$Centre de Nanosciences et de Nanotechnologies, CNRS, Univ. Paris-Sud, Universit\'{e} Paris-Saclay, F-91405 Orsay, France.\\
  ${}^2$CEA, INAC-PHELIQS, Nanophysique et semiconducteurs group, Univ. Grenoble Alpes, F-38000 Grenoble, France.\\
  ${}^3$Laboratoire Charles Coulomb (L2C), Universit\'e de Montpellier, CNRS, F-34095 Montpellier, France.\\
  ${}^4$Universit\'{e} C\^{o}te d'Azur, CRHEA-CNRS, F-06560 Valbonne, France.\\}
   ${}^{a)}$Current address: Universit\'e Paris Diderot-Paris7, F-75013 Paris, France.\\
  ${}^{b)}$Electronic address: philippe.boucaud@crhea.cnrs.fr\\
\end{center}

\setcounter{equation}{0}
\setcounter{figure}{0}
\setcounter{table}{0}
\setcounter{page}{1}
\renewcommand{\theequation}{S\arabic{equation}}
\renewcommand{\thefigure}{S\arabic{figure}}
\renewcommand{\bibnumfmt}[1]{[S#1]}
\renewcommand{\citenumfont}[1]{S#1}

\maketitle

\section*{FDTD simulations}

\begin{figure}
\includegraphics[width=0.5\linewidth]{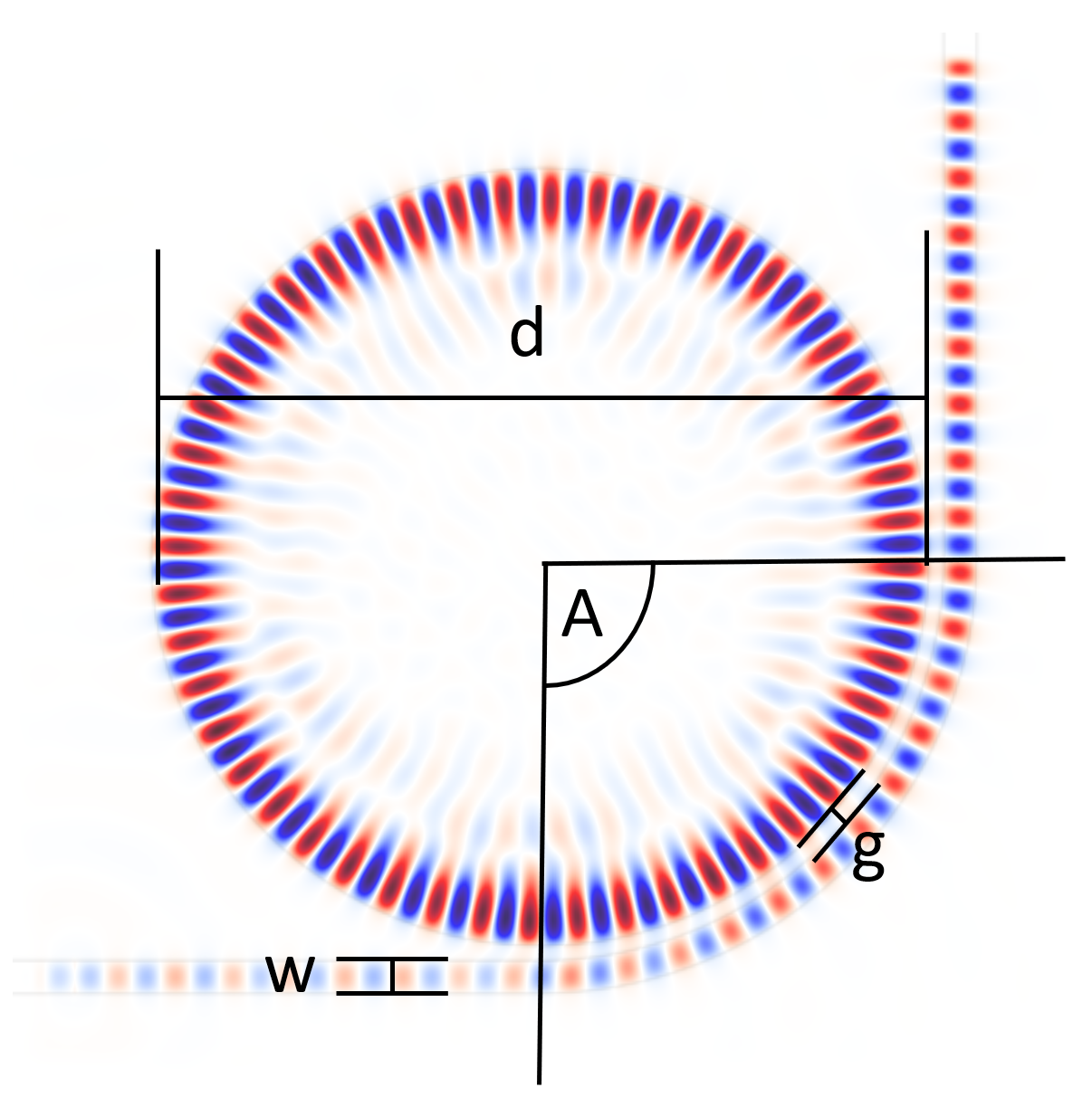}
  \caption{Simulation of the $H_z$ field of a $d=3~\mu \text{m}$ diameter disk for a first order radial mode with azimuthal order $m = 48$. The waveguide is bent at a $A=90^\circ$ angle, the gap is $g=50 ~\text{nm}$ and the waveguide width is $w =133 ~\text{nm}$.}
  \label{fig:fdtd}
\end{figure}

We have performed 3D finite-difference time-domain (FDTD) simulations for our fabricated microdisk photonic circuits using a resolution of 16 pixels per wavelength in the material. The simulated disk has a $3~\mu \text{m}$ diameter, an intrinsic Q factor of $Q_{int}=3600$, and a bus waveguide with a $90^\circ$ bending angle. The Q factor is limited by artificial absorption introduced in the disk to make the values more realistic. The structure consists of a two layer stack of 100 nm AlN ($n=2.12$) and 410 or 310 nm of GaN ($n=2.46$) for the disk and waveguide, respectively. A 2D in-plane visualization of a simulated device with a gap of $g=50 ~\text{nm}$ and a waveguide width of $w =133 ~\text{nm}$ is shown in figure \ref{fig:fdtd}, depicting the $H_z$ field of a first order radial mode with azimuthal order $m=48$.

\begin{figure}
\includegraphics[width=0.5\linewidth]{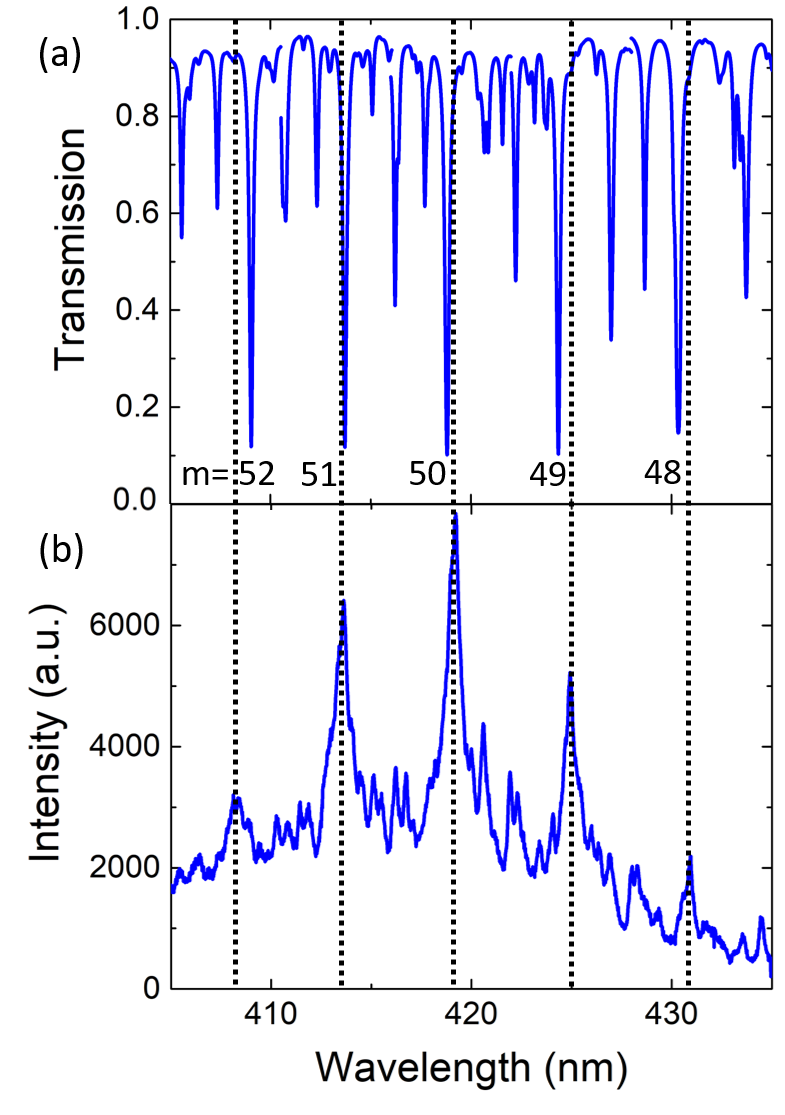}
  \caption{Comparison of a) simulation and b) measurement for a $3~\mu \text{m}$ diameter disk. The mode positions match well and the azimuthal numbers of the first order radial modes is clearly identified.}
  \label{fig:mode}
\end{figure}

In the simulation the disk is passive and we inject a Gaussian beam on the left side of the waveguide. We determine the transmission by taking the ratio between the flux at the end of the coupling waveguide and dividing it by the flux at the end of a straight uncoupled waveguide of the same width. The transmission formula is given by

\begin{equation}
T = \bigg(\frac{1-\frac{Q_{int}}{Q_c}}{1+\frac{Q_{int}}{Q_c}}\bigg)^2,
\end{equation}

where $Q_{int}$ is the intrinsic quality factor and $Q_c$ is the coupling quality factor.  A transmission spectrum is depicted in figure \ref{fig:mode} a) and compared to an experimental emission spectrum in figure \ref{fig:mode} b). The mode spacing between first order radial modes is $5-6~\text{nm}$ and the mode positions match well between simulation and experiment. The azimuthal number of each mode is clearly identified by performing simulations for a 1 nm wavelength range around each mode and counting the nodes in the $H_z$ field.

\begin{figure}
\includegraphics[width=0.8\linewidth]{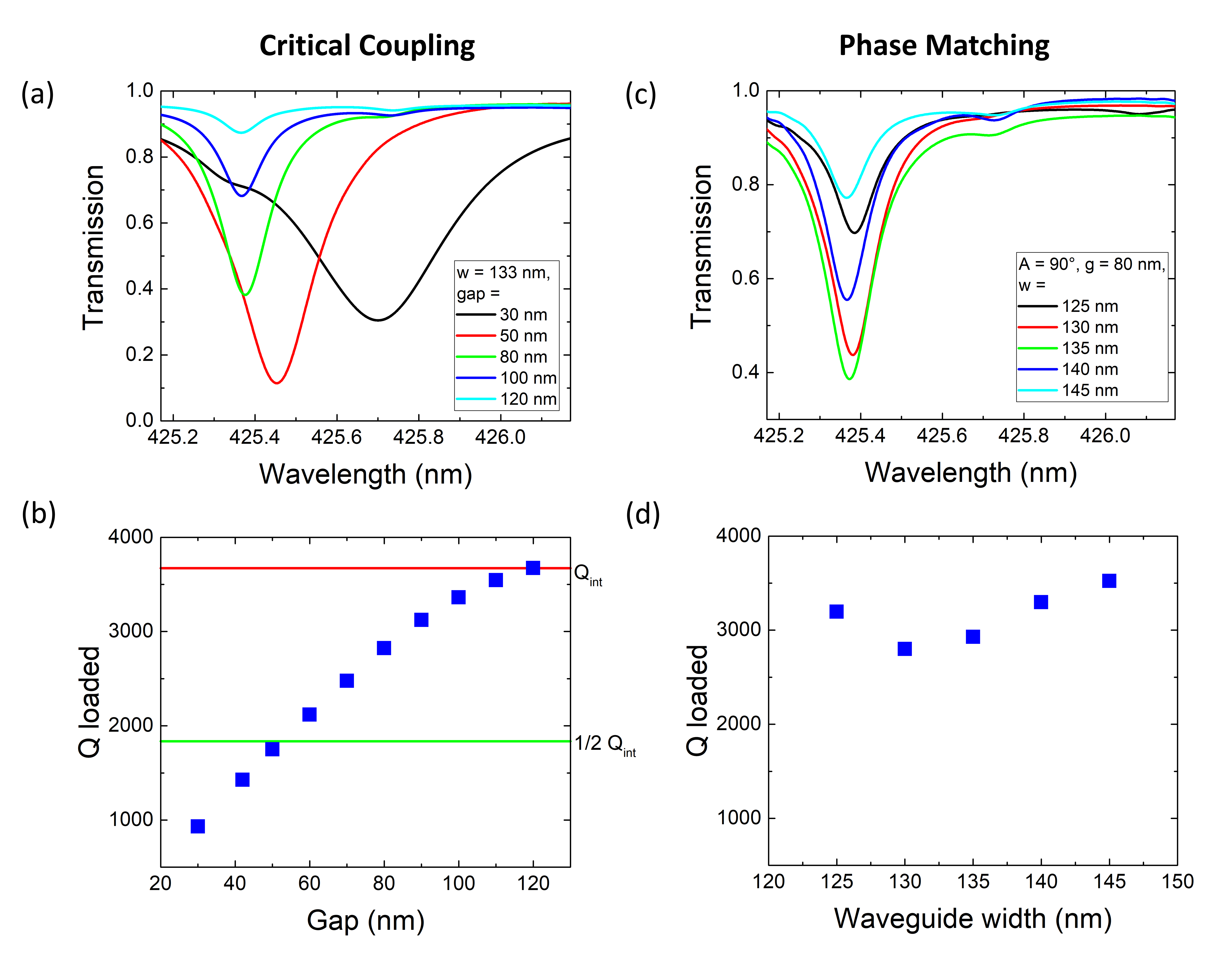}
  \caption{FDTD simulations of critical coupling and phase matching showing transmission and loaded Q factor of devices with a $3~\mu \text{m}$ diameter disk with $A=90^\circ$ for different gap sizes and waveguide widths a) Transmission for devices with $w=133~\text{nm}$ and five different gaps showing the undercoupled (80, 100, and 120 nm), critically coupled (50 nm) and over coupled (30 nm) regime. b) Loaded Q factor in dependence of gap size showing the degradation of the Q factor with increased coupling. $Q_{loaded}\approx Q_{int}$ for $g=120~\text{nm}$ and $Q_{loaded}\approx 1/2~Q_{int}$ at $g=50~\text{nm}$. c) Transmission for devices with a gap of 80 nm and different waveguide widths showing phase matching around $w=135$ nm. d) Loaded Q factor for different waveguide widths demonstrating its independence from phase matching far away from critical coupling.}
  \label{fig:crit}
\end{figure}

In figure \ref{fig:crit} a) the transmission spectra for one first order radial mode are depicted for different gap sizes. For a gap of 80, 100, and 120 nm the system is undercoupled, at 50 nm it is very close to critical coupling and a transmission minimum, and at 30 nm it is overcoupled. Figure \ref{fig:crit} b) shows the loaded Q factor over the gap size for gaps from 30 to 120 nm. For large gaps $Q_{loaded}\approx Q_{int}$. The loaded Q factor decreases with decreasing gap size and reaches $Q_{loaded}\approx 1/2~Q_{int}$ at critical coupling around $g=50~\text{nm}$. A further requirement for critical coupling is phase matching, which can be achieved for different combinations of $g$ and $w$ and is given at $g=50~\text{nm}~$ and $w= 133 ~\text{nm}$. Figure \ref{fig:crit} c) shows simulated transmission spectra for devices with a gap of 80 nm and different waveguide widths. For a width of around 135 nm a transmission minimum can be observed, which corresponds to phase matching. The corresponding loaded Q factor (depicted in figure \ref{fig:crit} d)) is not significantly degraded by phase matching when the system is far away from critical coupling, as is the case for a gap of 80 nm. Meanwhile we can observe the very high sensitivity of this system in the blue spectral range, as a small deviation of few tens of nm very rapidly degrade the transmission, i.e. the coupling ratio and visibility of the modes.

The simulation were performed for devices with a bending angle of $90^\circ$ but the results are similar at smaller angles and for straight waveguides. With decreasing waveguide bending angle the phase matching condition becomes less critical due to the reduced coupling length. Experimentally we are far away from critical coupling with gap sizes between 80 and 120 nm. Consequently, there is only a small degradation of the loaded Q factor, which is close to the intrinsic one. In this undercoupled regime the effect of phase matching on the Q factor is also minor, which allows us to observe lasing.

\begin{figure}
\includegraphics[width=1\linewidth]{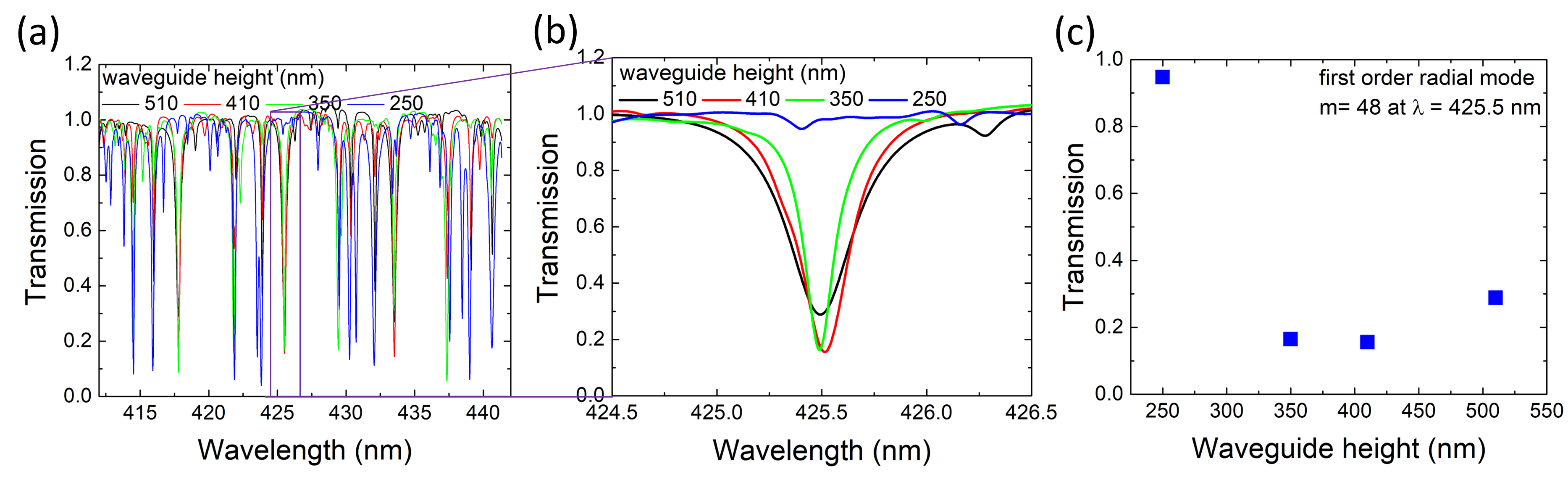}
  \caption{FDTD simulation of the influence of the waveguide height on the coupling strength. a) Transmission over a 30 nm wavelength range for a waveguide height between 250 nm and 510 nm. b) Zoom-in on the first-order radial mode with azimuthal order $m = 48$ at $425.5$ nm. c) Transmission over waveguide height for the same mode as in b).}
  \label{fig:wgheight}
\end{figure}

Figure \ref{fig:wgheight} a) shows FDTD simulations of devices with different waveguide heights between 250 nm and 510 nm over a 30 nm wavelength range, depicting various modes. A zoom-in onto the mode at 425.5 nm is shown in figure \ref{fig:wgheight} b). With a slight decrease in waveguide thickness as compared to the disk thickness, the coupling strength increases and a larger transmission dip is observed, while the mode position shifts for very thin waveguides. The transmission over the waveguide height for the mode at 425.5 nm is shown in figure \ref{fig:wgheight} c). The transmission minimum is observed at a waveguide height of 410 nm, which is close to our nominal etch depth.

\begin{figure}
\includegraphics[width=0.5\linewidth]{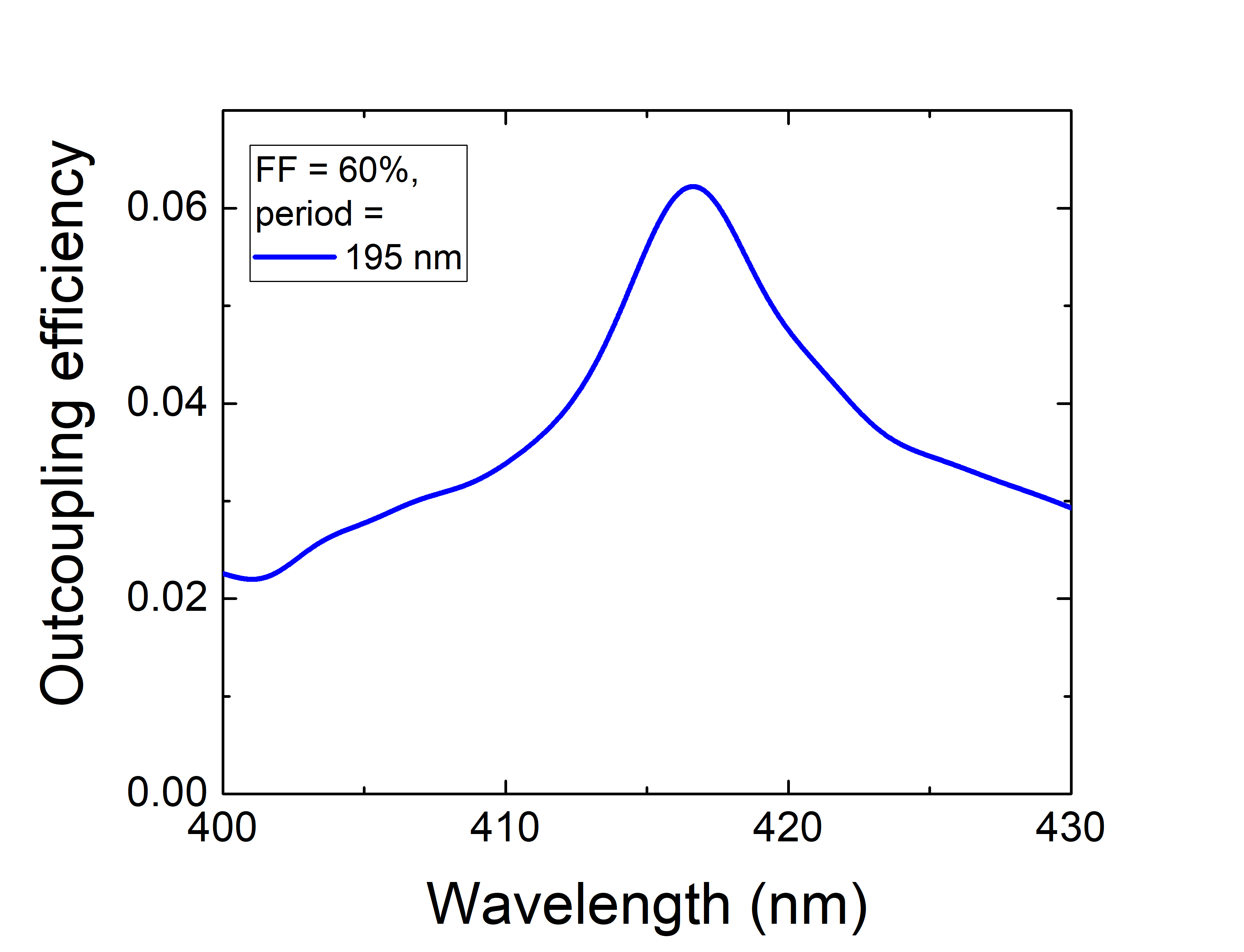}
  \caption{FDTD simulation of an out-coupling grating with a 195 nm period and a fill-factor of 60\%, giving an outcoupling efficiency of 6\% at 417 nm.}
  \label{fig:outcoupler}
\end{figure}

An FDTD simulation of an outcoupling grating is shown in figure \ref{fig:outcoupler} with a period of 195 nm and a fill-factor of 60\%. The outcoupling efficiency is 6\% at 417 nm.

\section*{Additional micro-PL results}

\begin{figure}
\includegraphics[width=1\linewidth]{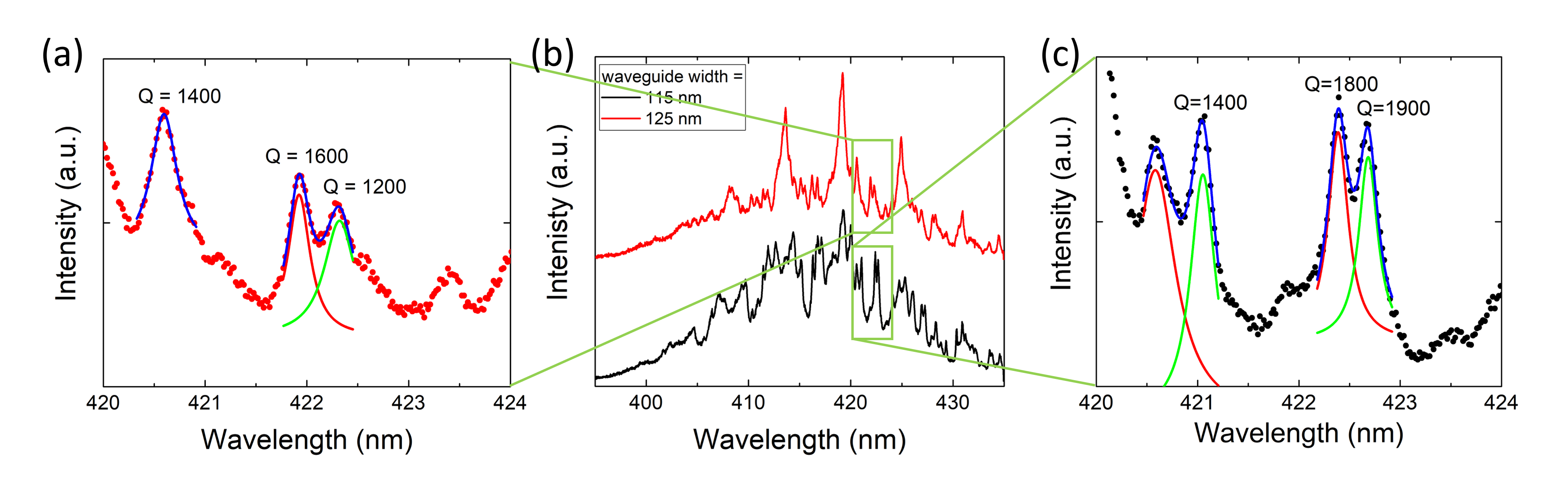}
  \caption{Loaded Q factor of two devices with a disk diameter of $3~\mu\text{m}$, a waveguide bending angle of $90^\circ$, a gap of 90 nm, and waveguide widths of 115 to 125 nm. a) Zoom-in of the $w=125~\text{nm}$ curve and Lorentzian fits giving loaded Q factors of 1200 to 1600. b) Full spectra of both devices. c) Zoom-in of the $w=115~\text{nm}$ curve and Lorentzian fits giving loaded Q factors of 1400 to 1900.}
  \label{fig:3um}
\end{figure}

Spectra of devices with $3~\mu\text{m}$ diameter disks, $90^\circ$ bent waveguides and a gap of 90 nm are shown in figure \ref{fig:3um} b) for waveguide widths of 115 nm and 125 nm. Zoom-ins of both curves are shown in figure \ref{fig:3um} a) and c), respectively, depicting Lorentzian fits of several modes, giving loaded Q factors between 1200 and 1900, which are very comparable to devices with $5~\mu\text{m}$ diameter, shown in figure 3 in the manuscript.

\begin{figure}
\includegraphics[width=0.5\linewidth]{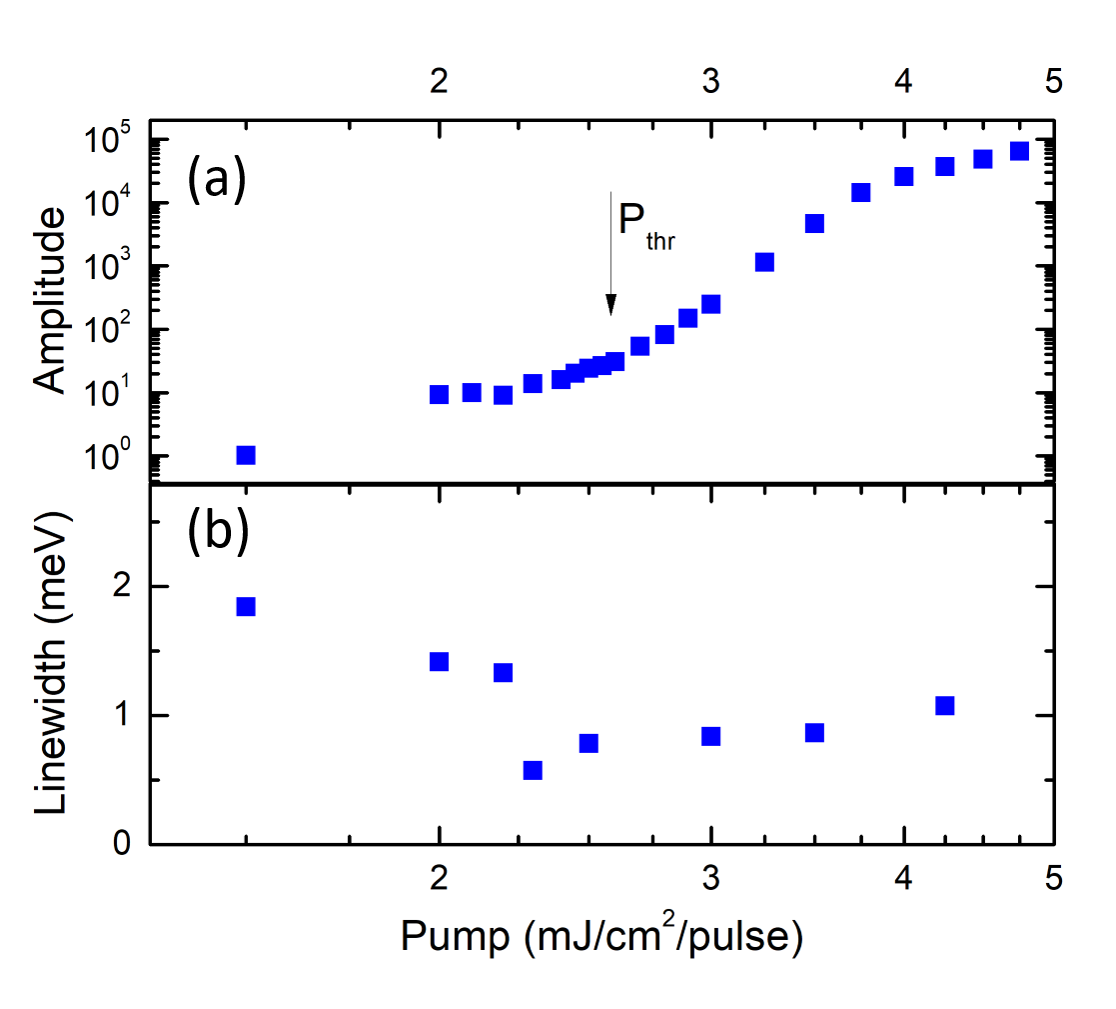}
  \caption{Pump energy-dependent measurement of a  $4~\mu\text{m}$ diameter microdisk laser without a waveguide fabricated on the same wafer, showing a) amplitude and b) linewidth of the lasing mode at 416.5 nm.}
  \label{fig:th}
\end{figure}

Using the same setup as described in the manuscript, we performed pulsed measurements for a $4~\mu\text{m}$ diameter disk without a waveguide with a lasing wavelength of 416.5 nm and a threshold of $2.6 ~\text{mJ/cm}^2$ per pulse. Figure \ref{fig:th} a) shows an S-shaped increase of the mode amplitude with pump energy and linewidth narrowing is observed in b).

\begin{figure}
\includegraphics[width=0.8\linewidth]{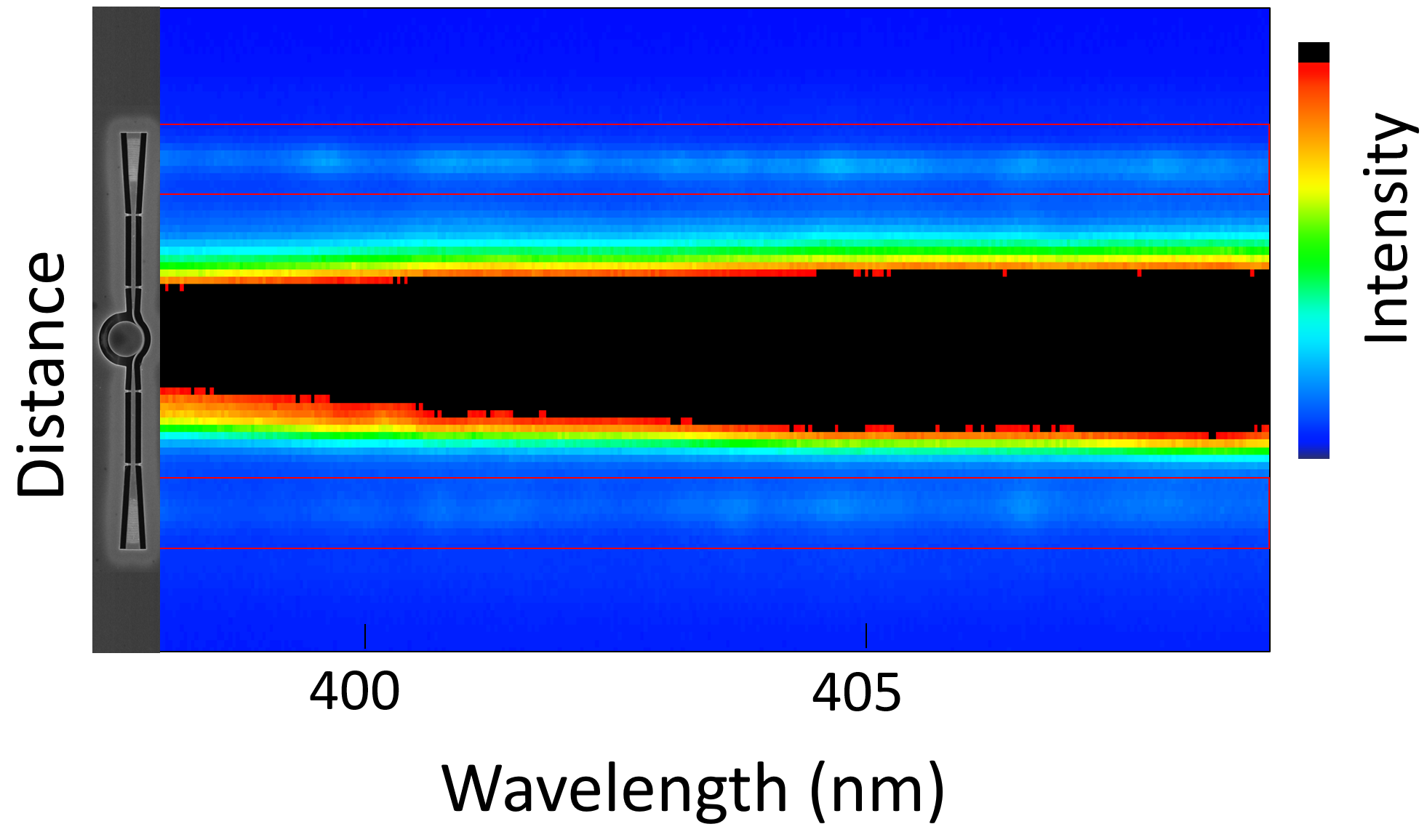}
  \caption{Over-saturated intensity map of the CCD with corresponding SEM image of a device on the left side. Same image as figure 2 a) in the manuscript, but with a different color scale to highlight the emission at the gratings. The gratings are indicated in red.}
  \label{fig:ccd}
\end{figure}

Figure \ref{fig:ccd} shows the CCD intensity map and corresponding SEM image of a device, also depicted in figure 2 a) in the manuscript with higher over-saturation of the CCD map to highlight the emission at the gratings, which is marked in red.

\section*{Additional microscope and SEM images}

\begin{figure}
\includegraphics[width=0.7\linewidth]{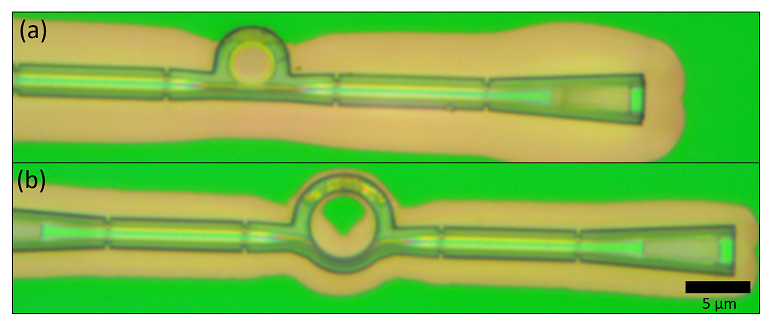}
  \caption{Optical microscope images of a) a $3~\mu\text{m}$ diameter disk with a straight waveguide and b) a $5~\mu\text{m}$ diameter disk with a $90^\circ$ bent waveguide. The underetch and silicon pedestal are visible.}
  \label{fig:micro}
\end{figure}

The pedestal shape varies based on the device geometry, due to the first ICP etch leaving a thin layer of AlN remaining. For devices with a $3~\mu\text{m}$ diameter disk (shown in figure \ref{fig:micro} a)), there is no pedestal remaining after the underetch and the disk is being held in place by the thin AlN layer. This explains the higher threshold and worse thermal management observed as compared to conventional mushroom-type microdisks. For  a $5~\mu\text{m}$ diameter disk and a $90^\circ$ waveguide angle (as shown in figure \ref{fig:micro} b))  we observe a triangular decentralized pedestal that could explain that lasing is not observed for such devices, due to mode leakage into the pedestal.

Close-up SEM images of various components of the photonic circuit are shown in figure \ref{fig:sem} a)-d). All components are showing some roughness from the processing.

\begin{figure}
\includegraphics[width=0.7\linewidth]{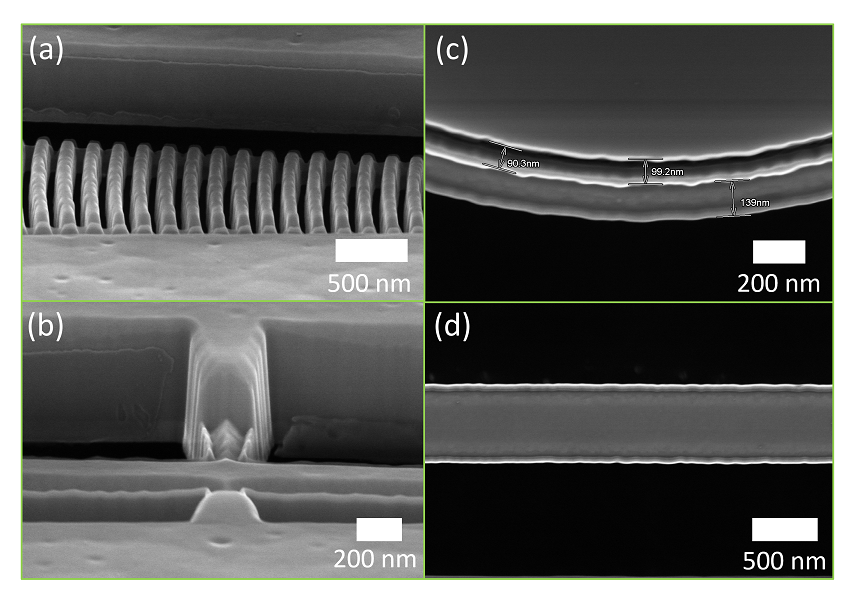}
  \caption{Close-up SEM images of various parts of the photonic circuit. a) Side-view of a grating coupler at a $20^\circ$ angle. b) Side-view of a tether and waveguide at a $20^\circ$ angle. c) Top-view of a disk and waveguide depicting a gap of 90 - 100 nm. d) Top-view of the wide part of a waveguide.}
  \label{fig:sem}
\end{figure}

\end{document}